\documentstyle[epsfig,a4p,12pt]{article}

\begin{document}


\begin{titlepage}
\begin{center}{\large  EUROPEAN LABORATORY FOR PARTICLE PHYSICS
}\end{center}\bigskip
\begin{flushright} CERN-PPE/97-123\\ 9 September 1997
\end{flushright}
\bigskip\bigskip\bigskip\bigskip\bigskip
\begin{center}{\huge\bf\boldmath
    Search for Unstable Heavy and Excited Leptons in e$^+$e$^-$ Collisions
    at $\sqrt{s} =$~170-172~GeV
}\end{center}\bigskip\bigskip
\begin{center}{\LARGE The OPAL Collaboration
}\end{center}\bigskip\bigskip

\begin{abstract}
We have searched for unstable neutral and charged
heavy leptons, N and ${\rm L^\pm}$, and
for excited states of neutral and charged leptons,
$\nu^*$, ${\rm e}^*$, $\mu^*$ and $\tau^*$,
in ${\rm e^+e^-}$ collisions at 
centre-of-mass energies of 
170 and 172~GeV using the
OPAL detector at LEP.  No evidence for their existence was found.
From the analysis of charged-current decays of pair-produced
unstable heavy leptons, and of charged-current and photonic decays
of pair-produced excited leptons, lower limits on their masses
are derived.
From the analysis of charged-current and photonic decays
of singly-produced excited leptons,
upper limits on
the ratio of the coupling to the compositeness scale,
$f/\Lambda$, are determined for masses up to
the kinematic limit.
\end{abstract}

%
%

\vspace{4cm}
\centerline{Submitted to Z. Phys. C}

\end{titlepage}
\begin{center}{\Large        The OPAL Collaboration
}\end{center}\bigskip
\begin{center}{
K.\thinspace Ackerstaff$^{  8}$,
G.\thinspace Alexander$^{ 23}$,
J.\thinspace Allison$^{ 16}$,
N.\thinspace Altekamp$^{  5}$,
K.J.\thinspace Anderson$^{  9}$,
S.\thinspace Anderson$^{ 12}$,
S.\thinspace Arcelli$^{  2}$,
S.\thinspace Asai$^{ 24}$,
D.\thinspace Axen$^{ 29}$,
G.\thinspace Azuelos$^{ 18,  a}$
% ...
}\end{center}\bigskip
\bigskip
% ...
$^{  2}$Dipartimento di Fisica dell' Universit\`a di Bologna and INFN,
I-40126 Bologna, Italy
\newline
% ...
$^{  5}$Cavendish Laboratory, Cambridge CB3 0HE, UK
\newline
% ...
$^{  8}$CERN, European Organisation for Particle Physics,
CH-1211 Geneva 23, Switzerland
\newline
$^{  9}$Enrico Fermi Institute and Department of Physics,
University of Chicago, Chicago IL 60637, USA
\newline
% ...
$^{ 12}$Indiana University, Department of Physics,
Swain Hall West 117, Bloomington IN 47405, USA
\newline
% ...
$^{ 16}$Department of Physics, Schuster Laboratory, The University,
Manchester M13 9PL, UK
\newline
% ...
$^{ 18}$Laboratoire de Physique Nucl\'eaire, Universit\'e de Montr\'eal,
Montr\'eal, Quebec H3C 3J7, Canada
\newline
% ...
$^{ 23}$Department of Physics and Astronomy, Tel Aviv University,
Tel Aviv 69978, Israel
\newline
$^{ 24}$International Centre for Elementary Particle Physics and
Department of Physics, University of Tokyo, Tokyo 113, and
Kobe University, Kobe 657, Japan
\newline
% ...
$^{ 29}$University of British Columbia, Department of Physics,
Vancouver BC V6T 1Z1, Canada
\newline
% ...
$^{  a}$ and at TRIUMF, Vancouver, Canada V6T 2A3
\newline
% ...
\newline


\newpage


\section{Introduction}
\label{s:intro}

In spite of its remarkable success in describing 
all electroweak data available today, the Standard Model
leaves many questions unanswered. In particular, it does not explain
the number of fermion generations nor the fermion mass spectrum.
The precise measurements of the Z-boson parameters
in ${\rm e^+e^-}$ collisions at centre-of-mass energies,
$\sqrt{s}$, around the Z-boson mass, $M_{\rm Z}$, have
determined the number of species of light neutrinos to be 
three~\cite{ref:pdg};
however, this does not exclude a fourth generation,
or other more exotic massive fermions,
if all of the new particles
have masses greater than $M_{\rm Z}/2$.
New fermions could be of the following 
types (for a review see Reference~\cite{ref:revue}):
(i) sequential fermions,
(ii) mirror fermions (with chirality opposite to those of the Standard Model),
(iii) vector fermions (with left- and right-handed doublets) and
(iv) singlet fermions.
These new fermions could be produced at high energy ${\rm e^+e^-}$ colliders
such as LEP,
where two production mechanisms are possible: (i) pair-production, 
and (ii) single-production in association with a light standard
fermion from the three known generations.
We concentrate on the search for the pair-production of
new unstable heavy leptons and for both pair- and single-production
of excited states of the known leptons
in ${\rm e^+e^-}$ collisions at centre-of-mass
energies up to $\sqrt{s}=$~172.3~GeV.
The results will be interpreted in terms of sequential and
excited lepton models.

Lower limits on the masses of heavy leptons were obtained 
at $\sqrt{s} \sim M_{\rm Z}$ \cite{ref:pdg,ref:hllep1},
and recent searches at $\sqrt{s}=$~130--140~GeV \cite{ref:hlOPAL15,ref:hllep15}
and $\sqrt{s} =$~161~GeV \cite{ref:hlOPAL161}
have improved these limits.
The L3 Collaboration has also reported results on 
sequential heavy lepton searches up to $\sqrt{s} =$~172~GeV
\cite{ref:hlL3172}.
Excited leptons have been sought at the LEP ${\rm e^+e^-}$
collider at $\sqrt{s} \sim M_{\rm Z}$
\cite{ref:ellep1},
$\sqrt{s}=$~130-140~GeV
\cite{ref:elopal15,ref:ellep15} and
$\sqrt{s}=$~161~GeV \cite{ref:elopal161,ref:ellep161},
and at the HERA ep collider \cite{ref:herasearches}.
Processes such as ${\rm e^+e^-} \rightarrow \gamma\gamma$ and
${\rm e^+e^-} \rightarrow {\rm f \bar{f}}$
are sensitive to new particles at higher mass scales
but have less sensitivity than direct searches if direct production
is kinematically allowed \cite{ref:opalgg,ref:2f}.


\subsection{Heavy Leptons}
\label{ss:hlintro}

We assume mixing with at most
one of the light standard leptonic generations
in order to avoid leptonic flavour-changing neutral currents (FCNC) at
the one-loop level.
General new heavy leptons could in principle decay
through the charge-current (CC) or the neutral-current (NC) channels:
\[
  {\rm N \rightarrow \ell^\pm W^\mp\quad\quad,\quad\quad
       N \rightarrow L^\pm W^\mp\quad\quad,\quad\quad
       N \rightarrow \nu_\ell Z},
\]
\[
  {\rm L^\pm \rightarrow \nu_\ell W^\pm \quad\quad ,\quad\quad
       L^\pm \rightarrow N W^\pm \quad\quad,\quad\quad
       L^\pm \rightarrow \ell^\pm Z},
\]
where N is a neutral heavy lepton, ${\rm L^\pm}$ is a charged heavy
lepton, and $\, {\rm \ell = e,\ \mu\ or\ \tau}$.
For heavy lepton masses less than the boson mass, $M_{\rm W}$ or $M_{\rm Z}$, 
the vector bosons are virtual, leading to a 3-body decay topology.
For masses greater than $M_{\rm W}$ or $M_{\rm Z}$, the decays are 2-body, and
the CC and NC branching ratios can be comparable.
For masses close to $M_{\rm W}$ or $M_{\rm Z}$, it is important to treat the
transition from the 3-body to the 2-body decay properly, including
effects from the vector boson widths.
Expressions for the computation of partial decay widths with an off-shell
W or Z can be found in~\cite{ref:abdel}.
A mixing angle ($\zeta$) of the heavy lepton with the standard lepton
flavour of 0.01 yields a decay length $c\tau$ of $\cal{O}$(1 nm);
since the decay length is proportional to $1/\zeta^2$,
by looking for unstable heavy leptons with efficiency for decays with
radii $\cal{O}$(1 cm), we remain sensitive down to
mixing angles squared $\cal{O}$(10$^{-12}$).
The present upper limit on
$\zeta^2$ is approximately 0.005~\cite{ref:nardi}.

Furthermore, we limit ourselves to the case where N and ${\rm L^\pm}$
decay via the CC channel only, as would be expected in a naive fourth
generation extension to the Standard Model.  In fact,
since the new mass region to which we are most sensitive is
near $M_{\rm W}$, the CC channel should dominate even for more exotic heavy leptons.
 
There are several possible new heavy lepton discovery channels,
depending on the charge of the lightest new heavy lepton and its decay modes.
We conduct three different analyses to ensure coverage of
these channels:\\ [10pt]
\noindent (A) ${\rm e^+e^-}\rightarrow{\rm N\bar{N}}$
with a flavour-mixing decay into a light lepton,
${\rm N \rightarrow e W}$, ${\rm N \rightarrow \mu W}$
  or ${\rm N\rightarrow \tau W}$.
  For a Dirac- and a Majorana- ($\mathrm{N=\bar{N}}$)
 type of neutral heavy lepton:
\begin{center}
\vbox{
\underline{Dirac}\hskip 6.5cm\underline{Majorana}\\[11pt]
\hskip -2cm
\begin{picture}(120,100)
  \put(10,90){$\mathrm{e^+e^-\longrightarrow N\, \bar{N}}$}
  \put(76,74){$\lfloor$}
  \put(79,68.2){$\mathrm{\rightarrow \ell^+ W^-}$}
  \put(112,55){$\lfloor$} 
  \put(115,49.2){$\mathrm{\rightarrow} q_i\bar{q}_j\ or\ \ell^{\prime -}\bar\nu_{\ell^\prime}$}
  \put(65,82){\line(0,-1){50}}
  \put(62.5,32){$\lfloor$}
  \put(65.5,26.2){$\mathrm{\rightarrow \ell^- W^+}$}
  \put(98.5,13){$\lfloor$} 
  \put(101.5,7.3){$\mathrm{\rightarrow} q_k\bar{q}_m\ or\ \ell^{\prime\prime +}\nu_{\ell^{\prime\prime}}$}
\end{picture}
\hskip 3cm
\begin{picture}(120,100)
  \put(10,90){$\mathrm{e^+e^-\longrightarrow N\, \bar{N}}$}
  \put(76,74){$\lfloor$}
  \put(79,68.2){$\mathrm{\rightarrow \ell^\pm W^\mp}$}
  \put(112,55){$\lfloor$} 
  \put(115,49.2){$\mathrm{\rightarrow} q_i\bar{q}_j\ or\ \ell^{\prime \mp}
\stackrel{_{_{(-)}}}{\nu_{\ell^\prime}}$}
  \put(65,82){\line(0,-1){50}}
  \put(62.5,32){$\lfloor$}
  \put(65.5,26.2){$\mathrm{\rightarrow \ell^\pm W^\mp}$}
  \put(98.5,13){$\lfloor$} 
  \put(101.5,7.3){$\mathrm{\rightarrow} q_k\bar{q}_m\ or\ \ell^{\prime\prime\mp}
\stackrel{_{_{(-)}}}{\nu_{\ell^{\prime\prime}}}$}
\end{picture}
}
\end{center}
The typical signature of such an event is at least two isolated charged
leptons plus the decay products of the two W bosons.  
In the Majorana case, the two leptons may have the same charge.
This should be the first visible channel if N is the lightest new
particle and it mixes with one of the known lepton generations.\\ [12pt]
\noindent (B) ${\rm e^+e^-}\rightarrow{\rm L^+L^-}$
with a flavour-mixing decay into a light neutrino
${\rm L^- \rightarrow \nu_\ell W^-}$:
\begin{center}
\hskip -2.0cm
\begin{picture}(120,100)
  \put(10,90){$\mathrm{e^+e^-\longrightarrow L^+\, L^-}$}
  \put(76,74){$\lfloor$}
  \put(79,68.2){$\mathrm{\rightarrow \nu_\ell W^-}$}
  \put(112,55){$\lfloor$} 
  \put(115,49.2){$\mathrm{\rightarrow} q_i\bar{q}_j\ or\ \ell^{\prime -}\bar\nu_{\ell^\prime}$}
  \put(65,82){\line(0,-1){50}}
  \put(62.5,32){$\lfloor$}
  \put(65.5,26.2){$\mathrm{\rightarrow \bar{\nu}_\ell W^+}$}
  \put(98.5,13){$\lfloor$} 
  \put(101.5,7.3){$\mathrm{\rightarrow} q_k\bar{q}_m\ or\ \ell^{\prime\prime +}\nu_{\ell^{\prime\prime}}$}
\end{picture}
\end{center}
This process leads typically to multi-jet events with large transverse momentum or
transverse missing energy.
This should be the first visible channel if ${\rm L^-}$ is the
lightest new particle and it mixes with one of the known lepton
generations.\\ [12pt]
\noindent (C) ${\rm e^+e^-}\rightarrow{\rm L^+L^-}$
where ${\rm L^- \rightarrow N W^-}$ (where
     N is a stable or long-lived neutral heavy lepton which decays outside the detector). 
     This production is possible if N is the lighter member of
     an SU(2) doublet which does not mix with the lighter generations.
\begin{center}
\hskip -2.0cm
\begin{picture}(120,100)
  \put(10,90){$\mathrm{e^+e^-\longrightarrow L^+\, L^-}$}
  \put(76,74){$\lfloor$}
  \put(79,68.2){$\mathrm{\rightarrow N\, W^-}$}
  \put(112,55){$\lfloor$} 
  \put(115,49.2){$\mathrm{\rightarrow} q_i\bar{q}_j\, ,\ \ell^{\prime -}\bar\nu_{\ell^\prime}$}
  \put(65,82){\line(0,-1){50}}
  \put(62.5,32){$\lfloor$}
  \put(65.5,26.2){$\mathrm{\rightarrow \bar{N}\, W^+}$}
  \put(98.5,13){$\lfloor$} 
  \put(101.5,7.3){$\mathrm{\rightarrow} q_k\bar{q}_m\, ,\ \ell^{\prime\prime +}\nu_{\ell^{\prime\prime}}$}
\end{picture}
\end{center}
This signal leads to a very low visible energy if the mass difference,
$M_{\rm L^\pm} - M_{\rm N}$, is small.
This would be the signature of a naive fourth generation
lepton doublet which does not mix appreciably
with the three known lepton generations and 
which satisfies $M_{\rm L^\pm} > M_{\rm N}$.

For completeness, a final search should be performed looking for long-lived
charged heavy leptons, which could come from a fourth generation
with $M_{\rm L^\pm} < M_{\rm N}$ and very small mixing.  The signature
would be a pair
of heavy charged particles visible in the detector.
This topology will be included in a subsequent paper.


\subsection{Excited Leptons}
\label{ss:elintro}

Excited leptons are new massive states of leptons expected
in compositeness models.  They are assumed to have the
same electroweak
SU(2) and U(1) gauge couplings to the vector bosons,
$g$ and $g^\prime$, as the
Standard Model leptons, but are expected to be grouped into
both left- and right-handed weak isodoublets with
vector couplings.  
The existence of the right-handed doublets is required to protect the
ordinary light leptons from radiatively acquiring a large
anomalous magnetic moment via the $\ell^*\ell V$
interaction \cite{ref:bdk}
($V$ is a $\gamma$, Z or W$^\pm$ vector boson).

Excited leptons could be produced in pairs in ${\rm e^+e^-}$ collisions
via the process ${\rm e^+e^-} \rightarrow \ell^* \bar{\ell}^*$,
governed by the $\ell^*\ell^*V$ coupling.  Depending on
the details of the three $\ell^*\ell V$ couplings, they could be
detected in photonic, CC or NC channels:
\[
  {\rm \nu_\ell^*  \rightarrow \nu_\ell\gamma     \quad\quad , \quad\quad
       \nu_\ell^*  \rightarrow \ell^\pm W^\mp     \quad\quad , \quad\quad
       \nu_\ell^*  \rightarrow \nu_\ell Z,
  }
\]
\[
  {\rm \ell^{*\pm} \rightarrow \ell^\pm \gamma    \quad\quad , \quad\quad
       \ell^{*\pm} \rightarrow \nu_\ell  W^\pm     \quad\quad , \quad\quad
       \ell^{*\pm} \rightarrow \ell^\pm  Z,
  }
\]
where $\nu^*$ is a neutral excited lepton, $\ell^{*\pm}$ is a charged
excited lepton, and $\ell =$~e, $\mu$ or $\tau$.

The branching fractions of the excited leptons into the different
vector bosons is determined by the strength of the three
$\ell^*\ell V$ couplings, which also determine the size of
the single-production cross-section.
We use the effective Lagrangian \cite{ref:bdk}
\[
  {\cal L}_{\ell\ell^*} =
  \frac{1}{2 \Lambda} \bar{\ell}^*\sigma^{\mu\nu}
  \left[g f \frac{ \mbox{\boldmath $\tau$} }{2}
    \mbox{\boldmath {\rm\bf W}}_{\mu\nu} +
  g^\prime f^\prime \frac{Y}{2} B_{\mu\nu} \right] \ell_{\rm L} +
  {\rm hermitian~conjugate},
\]
which describes the generalized magnetic de-excitation of
the $\ell^*$ states.
The matrix
$\sigma^{\mu\nu}$ is the covariant bilinear tensor,
\mbox{\boldmath $\tau$} are the Pauli matrices,
${\rm\bf W}_{\mu\nu}$ 
and $B_{\mu\nu}$ represent the fully gauge-invariant
field tensors
and $Y$ is the weak hypercharge.
The parameter $\Lambda$ has units of energy and can be regarded as
the ``compositeness scale,'' while $f$ and $f^\prime$ are couplings
associated with the different gauge groups.

Depending on the relative values of $f$ and $f^\prime$,
at the mass scales within the sensitivity of LEP
either the photonic or CC decay tends to have the largest
branching fraction, and in this paper we will consider
only these two decay modes.  For simplicity, we will
interpret our results using two example complementary
coupling assignments, $f=f^\prime$ and $f=-f^\prime$.

Several analyses are used for the excited lepton
search:\\[12pt]
\noindent (A) ${\rm e^+e^-} \rightarrow \nu^*\bar{\nu}^{(*)}$,
with CC decays of the neutral excited lepton(s):
\begin{center}
\vbox{
\underline{Pair Production}\hskip 4.5cm\underline{Single Production}\\[11pt]
\hskip -2cm
\begin{picture}(120,100)
  \put(10,90){$\mathrm{e^+e^-\longrightarrow \nu_\ell^*\bar\nu_\ell^*}$}
  \put(76,74){$\lfloor$}
  \put(79,68.2){$\mathrm{\rightarrow \ell^+ W^-}$}
  \put(112,55){$\lfloor$} 
  \put(115,49.2){$\mathrm{\rightarrow} q_i\bar{q}_j\ or\ \ell^{\prime -}\bar\nu_{\ell^\prime}$}
  \put(65,82){\line(0,-1){50}}
  \put(62.5,32){$\lfloor$}
  \put(65.5,26.2){$\mathrm{\rightarrow \ell^- W^+}$}
  \put(98.5,13){$\lfloor$}
  \put(101.5,7.3){$\mathrm{\rightarrow} q_k\bar{q}_m\ or\ \ell^{\prime\prime +}\nu_{\ell^{\prime\prime}}$}
\end{picture}
\hskip 3cm
\begin{picture}(120,100)
  \put(10,90){$\mathrm{e^+e^-\longrightarrow \nu_\ell^*\nu_\ell}$}
  \put(62.5,74){$\lfloor$}
  \put(65.5,68.2){$\mathrm{\rightarrow \ell^\pm W^\mp}$}
  \put(98.5,55){$\lfloor$} 
  \put(101.5,49.2){$\mathrm{\rightarrow} q_i\bar{q}_j\ or\ \ell^{\prime\mp}
    \stackrel{_{_{(-)}}}{\nu_{\ell^{\prime}}}$}
\end{picture}
}
\end{center}
The topology of pair-production with CC decays is almost identical to
the flavour-mixing decay of neutral heavy leptons considered
as case~(A)
in Section~\ref{ss:hlintro}, and the same analysis is used.
The topology of single-production with CC decays can be categorised
by the W$^\pm$ decay mode.  If the W$^\pm$ decays leptonically,
the signature is two leptons with missing transverse momentum
(acoplanar lepton pair),
while if it decays hadronically, the signature is an isolated
lepton with two hadronic jets plus missing energy.
In this latter case, the mass of the excited lepton can
be reconstructed.\\[12pt]
\noindent (B) ${\rm e^+e^-} \rightarrow \ell^{*\pm}\ell^{(*)\mp}$,
with CC decays of the charged excited lepton(s):
\begin{center}
\vbox{
\underline{Pair Production}\hskip 4.5cm\underline{Single Production}\\[11pt]
\hskip -2cm
\begin{picture}(120,100)
  \put(10,90){$\mathrm{e^+e^-\longrightarrow \ell^{*+}\ell^{*-}}$}
  \put(76,74){$\lfloor$}
  \put(79,68.2){$\mathrm{\rightarrow \nu_\ell W^-}$}
  \put(112,55){$\lfloor$} 
  \put(115,49.2){$\mathrm{\rightarrow} q_i\bar{q}_j\ or\ \ell^{\prime -}\bar\nu_{\ell^\prime}$}
  \put(65,82){\line(0,-1){50}}
  \put(62.5,32){$\lfloor$}
  \put(65.5,26.2){$\mathrm{\rightarrow \bar{\nu}_\ell W^+}$}
  \put(98.5,13){$\lfloor$} 
  \put(101.5,7.3){$\mathrm{\rightarrow} q_k\bar{q}_m\ or\ \ell^{\prime\prime +}\nu_{\ell^{\prime\prime}}$}
\end{picture}
\hskip 3cm
\begin{picture}(120,100)
  \put(10,90){$\mathrm{e^+e^-\longrightarrow \ell^{*\pm}\ell^\mp}$}
  \put(62.5,74){$\lfloor$}
  \put(65.5,68.2){$\mathrm{\rightarrow
      \stackrel{_{_{(-)}}}{\nu_\ell}
      W^\pm}$}
  \put(98.5,55){$\lfloor$} 
  \put(101.5,49.2){$\mathrm{\rightarrow} q_i\bar{q}_j\ or\ \ell^{\prime\pm}
    \stackrel{_{_{(-)}}}{\nu_{\ell^\prime}}$}
\end{picture}
}
\end{center}
Again, the analysis for the flavour-mixing decay of charged
heavy leptons considered as case~(B)
in Section~\ref{ss:hlintro} is used for pair-production.
The topologies of single-production with CC decays are the same
as that for neutral excited leptons, and the same analysis
is used.\\[12pt]
\noindent (C) ${\rm e^+e^-} \rightarrow \nu^*\bar{\nu}^{(*)}$,
with photonic decays of the neutral excited lepton(s):
\begin{center}
\vbox{
\underline{Pair Production}\hskip 4.5cm\underline{Single Production}\\[11pt]
\hskip -2cm
\begin{picture}(120,60)
  \put(10,50){$\mathrm{e^+e^-\longrightarrow \nu^*\bar\nu^*}$}
  \put(76,34){$\lfloor$}
  \put(79,28.2){$\mathrm{\rightarrow \bar\nu \gamma}$}
  \put(65,42){\line(0,-1){30}}
  \put(62.5,12){$\lfloor$}
  \put(65.5,6.2){$\mathrm{\rightarrow \nu \gamma}$}
\end{picture}
\hskip 3cm
\begin{picture}(120,60)
  \put(10,50){$\mathrm{e^+e^-\longrightarrow \nu^*\nu}$}
  \put(62.5,34){$\lfloor$}
  \put(65.5,28.2){$\mathrm{\rightarrow \nu \gamma}$}
\end{picture}
}
\end{center}
The signature is one or two photons plus significant
missing energy.  While no direct mass reconstruction
is possible, the kinematics of the photon(s) can
be used to restrict the excited neutrino mass which is consistent
with the event.\\[12pt]
\noindent (D) ${\rm e^+e^-} \rightarrow \ell^{*\pm}\ell^{(*)\mp}$,
with photonic decays of the charged excited lepton(s):
\begin{center}
\vbox{
\underline{Pair Production}\hskip 4.5cm\underline{Single Production}\\[11pt]
\hskip -2cm
\begin{picture}(120,60)
  \put(10,50){$\mathrm{e^+e^-\longrightarrow \ell^{*+}\ell^{*-}}$}
  \put(76,34){$\lfloor$}
  \put(79,28.2){$\mathrm{\rightarrow \ell^- \gamma}$}
  \put(65,42){\line(0,-1){30}}
  \put(62.5,12){$\lfloor$}
  \put(65.5,6.2){$\mathrm{\rightarrow \ell^+ \gamma}$}
\end{picture}
\hskip 3cm
\begin{picture}(120,60)
  \put(10,50){$\mathrm{e^+e^-\longrightarrow \ell^{*\pm}\ell^\mp}$}
  \put(62.5,34){$\lfloor$}
  \put(65.5,28.2){$\mathrm{\rightarrow \ell^\pm \gamma}$}
\end{picture}
}
\end{center}
The signature is two leptons plus one or two photons with
no missing energy, and the excited lepton mass can
be reconstructed from the lepton-photon invariant mass.
If the ${\rm e^*e}\gamma$ coupling is significant,
excited electron single-production is dominated by
$t$-channel photon exchange; in this case, the recoil
electron is most often missing along the beam axis, making
a search in the ${\rm e}\gamma$ plus missing electron
topology interesting as well.


\section{The OPAL Detector and Data Sample}
\label{s:opaldet}

A complete description of the  OPAL detector can be found 
in Reference \cite{ref:OPAL-detector}, and it is described only
briefly here.
The central detector consists of a system of tracking chambers
that provides charged particle tracking over 96\% of the full solid 
angle\footnote
   {The OPAL coordinate system is defined so that the $z$-axis is in the
    direction of the electron beam and the $x$-axis
    points towards the centre of the LEP ring; 
    $\theta$ and $\phi$
    are the polar and azimuthal angles, defined relative to the
    $+z$- and $+x$-axes, respectively.}
inside a uniform 0.435~T magnetic field. It consists of a two-layer
silicon microstrip vertex detector, a high-precision vertex drift
chamber,
a large-volume jet chamber and a set of $z$ chambers that measures
the track coordinates along the beam direction.
The specific ionization
energy loss per unit path length in the jet chamber, ${\rm d}E/{\rm d}x$,
is used for particle identification.
A lead-glass electromagnetic
calorimeter located outside the magnet coil
covers the full azimuthal range with excellent hermeticity
in the polar angle range of $|\cos \theta |<0.82$ for the barrel
region and $0.81<|\cos \theta |<0.984$ for the endcap region.
In the region where the barrel and endcap calorimeters overlap,
photon and electron energy resolution is somewhat degraded
because of extra material in front of the calorimeters.
A set of wire-chambers in front of the electromagnetic calorimeter,
the presampler, is used to measure the shape of 
electromagnetic showers from electrons and photons that interact
in the magnet coil, and is used to aid electron identification.
The magnet return yoke is instrumented with streamer tubes
with cathode-strip readout for hadron calorimetry
and consists of barrel and endcap sections along with
pole-tip detectors that
together cover the region $|\cos \theta |<0.99$.
Muons are identified with the hadron calorimeter strips,
and with four layers of muon chambers which
cover the outside of the hadron calorimeter.
The gamma-catcher, forward-detector and silicon-tungsten
electromagnetic calorimeters 
complete the geometrical acceptance down to 24~mrad.
The silicon-tungsten calorimeter 
is used for the luminosity measurement.

The primary data sample used in this paper
has an integrated luminosity of
1.0~pb$^{-1}$ at $\sqrt{s}=$~170.3~GeV and 
9.3~pb$^{-1}$ at $\sqrt{s}=$~172.3~GeV and was
acquired by OPAL during the autumn 1996 LEP running period.
The results are combined with those obtained from
10.3~pb$^{-1}$ of data at $\sqrt{s}=$~161.3~GeV acquired during
the summer of 1996.  The luminosities used in the purely
photonic events selection are slightly different due to
small differences in the detector status requirements used
in those analyses.


\section{Monte Carlo Simulation}
\label{s:mc}

A new Monte Carlo generator, EXOTIC, has been used for
the simulation of ${\rm e^+e^-}\rightarrow {\rm N \bar N}$ and
${\rm e^+e^-} \rightarrow {\rm L^+L^-}$ events.
The code is based
on formulae given in references~\cite{ref:abdel,ref:zerwas}
and the matrix elements
include all spin correlations in the production and decay processes
of the heavy leptons, and a complete treatment of the transition
from 3-body to 2-body heavy fermion decays;
it uses JETSET~\cite{ref:jetset} for fragmentation
and hadronization of quarks. 
For ${\rm N \bar N}$,  and for ${\rm L^+L^-}$
with ${\rm L^- \rightarrow \nu_\ell W^-}$, production, Monte Carlo
samples of 2000 events each were generated for a set of masses
in the region of sensitivity of this analysis.
Different samples for Dirac- and Majorana-type heavy
neutral leptons were generated, taking into account the different
angular distributions in the two cases.
For the case where
${\rm L^- \rightarrow N W^-}$,
samples of 1000 events each 
were simulated at 30 points in the ($M_{\rm L}$,$M_{\rm N}$) plane
with $M_{\rm L}$ ranging
from 70 to 85 GeV and $M_{\rm N}$ from 40 to 80 GeV,
and with a mass difference $M_{\rm L} - M_{\rm N}$ larger than 5 GeV.

The excited lepton Monte Carlo samples were simulated with the existing
generators EXLP and LSTAR described in Reference~\cite{ref:elopal15}.
These generators use the formulae from Reference~\cite{ref:bdk}, but
without taking into account the spin correlations in the production
and decay of excited leptons.  The effect of spin correlations on the
efficiency for the detection of excited leptons has been studied
using analytic approximations and also
using EXOTIC, which supports excited leptons as well, 
and it was found to be negligible.
For the pair production channels, samples of 2000 events each were generated
for masses in the range from 65 to 85~GeV, and for the single
production case samples of 2000 events each
were generated for masses in the range from 80 to 170~GeV.

The backgrounds from Standard Model processes were studied with
a variety of Monte Carlo generators.  Two-fermion processes
were simulated with BHWIDE \cite{ref:bhwide} (large-angle Bhabha scattering),
TEEGG \cite{ref:teegg} ($t$-channel Bhabha scattering),
KORALZ \cite{ref:koralz} (muon- and tau-pair
production) and PYTHIA \cite{ref:jetset}
(${\rm e^+e^- \rightarrow Z/\gamma \rightarrow q \bar q}$).
These generators all include both initial- and final-state
radiation, which is particularly important for the radiative decay
analyses.

Hadronic two-photon processes of the form
${\rm e^+e^- \rightarrow e^+e^- q \bar q}$, with the final-state
electrons scattered at small angles, were simulated with
PYTHIA, HERWIG \cite{ref:herwig} and PHOJET \cite{ref:phojet}.
Leptonic two-photon processes were simulated with
Vermaseren \cite{ref:vermaseren}.
Other 4-fermion processes,
the most important of which is ${\rm W^+W^-}$ production,
were studied with PYTHIA, FERMISV \cite{ref:fermisv},
EXCALIBUR \cite{ref:excalibur} and grc4f \cite{ref:grc4f}.
FERMISV simulates neutral-current 4-fermion processes, which
can be combined with ${\rm W^+W^-}$ production samples generated
with PYTHIA. EXCALIBUR and grc4f include CC and
NC 4-fermion processes and all interference effects.
Different combinations of samples from the different
4-fermion generators yielding a complete estimate of the
background processes gave consistent predictions.

Finally, Standard Model processes with only photons
in the final state are an important background to the analysis of
excited neutral leptons with photonic decays.
The RADCOR \cite{ref:radcor} program was used to simulate the process
${\rm e^+e^-} \rightarrow \gamma \gamma (\gamma)$, and
both KORALZ and
NUNUGPV \cite{ref:nunugpv} were used to simulate the process
${\rm e^+e^-} \rightarrow \nu\overline{\nu} \gamma (\gamma)$.

All signal and background Monte Carlo samples were
processed through the full OPAL detector simulation \cite{ref:gopal},
and passed through the same analysis chain as the data.


\section{Selection Criteria}
\label{s:selection}

Since the searches presented in this paper involve many different
experimental topologies, a number of different analyses are 
used for the heavy and excited lepton searches.  These analyses
evolved independently and use slightly different criteria for
such details as track and cluster quality requirements and lepton identification
methods.  In Section~\ref{ss:pphl}, the three analyses for
the CC decays of pair-produced heavy and excited leptons are described.  
Next,
Section~\ref{ss:spelw} details
the search for the single production of neutral and charged 
excited leptons with CC decays.  
Section~\ref{ss:nelg}
summarizes the search for neutral excited leptons with photonic decays.
Finally, 
in Section~\ref{ss:celg}, the selection of charged excited leptons
with photonic decays is described.  


\subsection{Pair Production of Heavy Leptons and of Excited Leptons with CC Decays}
\label{ss:pphl}

All charged tracks and calorimeter clusters are
subjected to some quality criteria:
each track is required to have
a minimum momentum transverse to the beam axis,
$p_{\rm T}$, of 0.15~GeV, to have a minimum number of 20 hits
in the central tracking detector, 
and to point to the geometrical centre of the detector 
to within 2 cm and to pass within
25 cm of the origin in $z$ at the point of closest approach in $r\phi$;
each electromagnetic cluster in the barrel is required to have
a minimum energy of
0.1 GeV, while for the endcap regions
the minimum raw energy is 0.25~GeV;
hadronic clusters are required to have a minimum energy of 0.25~GeV.
In calculating the total visible energy and 
momentum of events, and of individual jets,
corrections are applied which reduce the effects
of the double-counting of energy from tracks and
clusters associated to them \cite{ref:gce}.
Jet reconstruction is done by using 
the $k_\bot$ (``Durham'')~\cite{ref:durham} jet-finding algorithm
with a resolution parameter $y_{\rm cut}$ of 0.004. For electron, muon, and tau 
identification the following procedures are applied:

\begin{description}
\item[electron:] a track is identified as an electron
candidate if at least one of the following two methods is satisfied.
\begin{enumerate}
\item A standard electron identification: each good charged track
 in the central detector, which satisfies all the following criteria, is
     considered as an electron candidate:
  \begin{itemize}
    \item $p> 3.0 $ GeV, where $p$ is the track momentum.
    \item $ 0.7 < E/p < 1.4$ , where $E$ is the energy of the electromagnetic
          cluster associated to the track.
    \item $ 8 < {\rm d}E/{\rm d}x < 13$ keV/cm, where
          ${\rm d}E/{\rm d}x$ is the ionization energy loss per unit path length
           measured in the central detector.
    \item the number of ${\rm d}E/{\rm d}x$ measurements is at least 20.
  \end{itemize}
\item The output of an artificial neural network designed to
        identify electrons~\cite{ref:NN} is required to exceed 0.90. 
   The network
   uses information mainly from the track momentum, ${\rm d}E/{\rm d}x$,
   the energy in the associated electromagnetic calorimeter cluster, the number
   of lead-glass blocks in the cluster, and the presampler signal associated with the track.
\end{enumerate}
     In each case there is an isolation requirement on the electron candidate
     using the total energy 
     in a cone of half-opening angle 15$^\circ$ surrounding 
     the electron, excluding the energy of the electromagnetic cluster 
     associated with the candidate. The electron is considered to be 
     isolated if this energy does not
     exceed 5 GeV.
\item[muon:] a track with $p> 3.0$ GeV is identified as a muon
 candidate if it is matched with track segments in the muon chambers. 
In regions not covered
by the muon chambers the identification is done by using a match with the
hadron calorimeter strips. We require that muon candidates be isolated 
as described above for electrons.
\vskip 0.4cm
\item[tau:] a tau lepton candidate is selected by either the electron
or muon identification (as above) for a 1-prong tau decay into an isolated lepton
 (e or $\mu$), or by jet reconstruction for a 1-prong or a 3-prong 
tau decay into hadrons.
A jet is identified as a tau if either of the following criteria were satisfied.
\vskip -.3cm
\begin{enumerate}
  \item for a 1-prong decay, all of the following conditions must be satisfied:
    \begin{itemize}
      \item it contained at least 1 track with $p>$~3.0~GeV.
      \item no track was found within a cone of half-opening angle
            15$^\circ$ surrounding the track.
      \item the momentum sum of the other tracks within the jet was less than
            1.0~GeV.
      \item the invariant mass of all tracks and clusters within the cone
             was less than 3.0~GeV. 
    \end{itemize}
  \item for a 3-prong decay, all of the following conditions must be satisfied:
    \begin{itemize}
      \item it contained exactly 3 tracks in a cone of half-opening angle
          15$^\circ$ around the jet axis.
      \item the momentum sum of the 3 tracks was greater than 3.0 GeV.
      \item the invariant mass of all tracks and clusters within the cone
            was less than 3.0 GeV.
    \end{itemize}
\end{enumerate}
\end{description}

\subsubsection{\boldmath Selection of ${\rm N \bar N}$ Candidates}
\label{sss:NN}

\begin{figure}[b]
\centerline{\hbox{
\epsfig{file=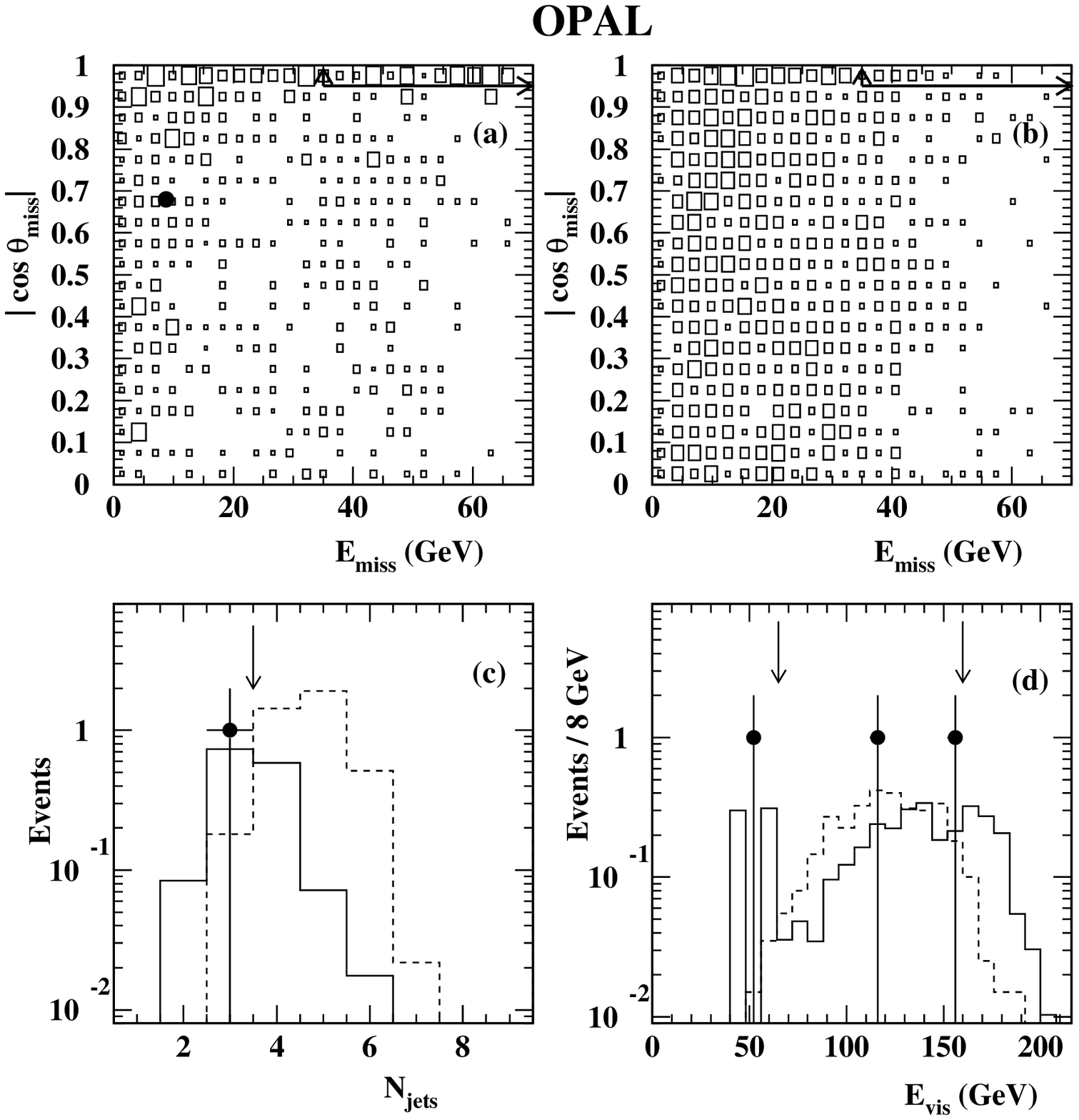,width=\textwidth}
}}
\caption{
  The distributions of $|\cos\theta_{\rm miss}|$ 
  versus $E_{\rm miss}$ are shown in (a) 
  for the background processes (box) and the selected 
  ${\rm N \rightarrow e W}$ event among the data (filled circle) 
  after cut~2; the distribution for the $\rm N\bar{N}$ signal
  is shown in (b) for a Dirac N (with $M_{\rm N}=75$ GeV).
  The distributions for the number of reconstructed jets 
  for the case ${\rm N \rightarrow e W}$ after cut~3 are
  shown in (c) for the data (filled circles), the background processes 
  (solid line), and the $\rm N\bar{N}$ signal 
  (Dirac N, with $M_{\rm N}=75$ GeV) (broken line).
  For the case ${\rm N \rightarrow \tau W}$ the  
  visible energy ($E_{\rm vis}$) distributions after cut 4
  are shown in (d) for the data (filled circles), the background processes 
  (solid line), and the $\rm N\bar{N}$ signal (broken line) 
  for a Dirac N with $M_{\rm N}=65$ GeV.
  The arrows correspond to the cut values.
  The corresponding distributions for a Majorana N are similar.
}
\label{f:NN}
\end{figure}

To be considered as potential candidates for ${\rm N\bar{N}}$,
or $\nu^*\bar\nu^*$ with CC decays
(category A in Section~\ref{ss:hlintro}, and pair-production
for category A in Section~\ref{ss:elintro}),
the events must pass the following set of cuts:

\begin{enumerate}
  \item A standard multihadronic selection where tracks and energy clusters 
        satisfy the following (good tracks and clusters are those which
        satisfy the quality criteria defined in Section~\ref{ss:pphl}):
        \begin{itemize}
          \item $R_{\rm vis}\ge 0.10$, where
                $R_{\rm vis}={E_{\rm shw}\over 2\times E_{\rm beam}}$;
                $E_{\rm shw}=\sum E_{\rm raw}$ and $E_{\rm raw}$ 
                is the raw energy of a good cluster
                and $E_{\rm beam}$ is the beam energy.
          \item $R_{\rm bal}\le 0.65$, where 
                $R_{\rm bal}={E_{\rm bal}\over E_{\rm shw}}$;
                $E_{\rm bal}=\sum E_{\rm raw}\cos\theta$ ($\theta$ is the
                polar angle of the cluster).
          \item there are at least 7 good clusters.
          \item there are at least 5 good tracks.
        \end{itemize} 
  \item There are at least 2 isolated leptons of the same flavour
        ($\rm e,\ \mu,\ or\ \tau$).
        We do not apply a charge constraint in order to be sensitive
        both to Dirac and Majorana neutral heavy leptons.
  \item To reject radiative events, principally
        ${\rm e^+e^-} \rightarrow {\rm Z}\gamma$ with the photon
        missing along the beam axis, we require that
        $|\cos\theta_{\rm miss}| < 0.95$ if $E_{\rm miss}> 35$ GeV, where
        $\theta_{\rm miss}$ is the angle
        between the missing energy vector and the beam axis
        and $E_{\rm miss}$ is the missing energy in the event
        (the most probable photon energy from ${\rm e^+e^-} \rightarrow {\rm Z}\gamma$
        is about 60~GeV).
        The distributions after cut~2 of $| \cos\theta_{\rm miss} |$
        versus $E_{\rm miss}$ are shown in Figures~\ref{f:NN}(a) and 
        \ref{f:NN}(b).
  \item The number of reconstructed jets is required to be at least four.
        The distributions after cut 3 of the numbers of
        reconstructed jets are shown in Figure~\ref{f:NN}(c).
  \item For the case ${\rm N \rightarrow \tau W}$ an extra cut is
        applied to reduce the larger background:
        we require that $65 <E_{\rm vis}< 160$ GeV, where $E_{\rm vis}$ is
        the visible energy. 
        The visible energy distributions after cut~4 
        are shown in Figure~\ref{f:NN}(d).
\end{enumerate}

For the cases ${\rm N \rightarrow e W}$ and ${\rm N \rightarrow \mu W}$
the remaining numbers of events after
each cut are displayed in
Tables~\ref{t:NNe} and~\ref{t:NNmu} respectively (numbers for Dirac- and 
Majorana-type N signals are shown in the same table).
The numbers of events remaining for ${\rm N \rightarrow \tau W}$
after each cut are displayed in Table~\ref{t:NNtau}.

The efficiencies obtained for the ${\rm N \bar N}$ and 
$\nu^*\bar\nu^*$
selections after applying
the full analysis are evaluated from the Monte Carlo samples
discussed in Section~\ref{s:mc}.
The efficiencies are about 60\% for
${\rm N \rightarrow e W}$, 50\% for ${\rm N \rightarrow \mu W}$,
and 30\% for ${\rm N \rightarrow \tau W}$.

\begin{table}
\begin{center}
\begin{tabular}{|c||c|c||c|c|c|c||c|c|} \hline
 After & Data & Total & ${\rm q \bar q}(\gamma)$ & two-photon &$\tau\tau(\gamma)$ &
 4-f   & Dirac & Majorana \\
 Cut   & & Bkg & & & & & $M_{\rm N}=75$ GeV & $M_{\rm N}=65$ GeV  \\ \hline
Presel &  $-$  &  $-$ &  $-$  & $-$  & $-$  &  $-$  & 6.26 & 4.59 \\
 1     &  1411 & 1288 &  1137 & 26.6 & 3.33 & 120.8 & 5.67 & 4.15 \\
 2     &  1    & 1.72 &  0.35 & 0.0  & 0.03 & 1.34  & 4.07 & 3.14 \\
 3     &  1    & 1.48 &  0.21 & 0.0  & 0.03 & 1.25  & 4.01 & 3.07 \\
 4     &  0    & 0.68 &  0.08 & 0.0  & 0.01 & 0.59  & 3.88 & 2.89 \\ \hline
\end{tabular}
\caption{Observed number of events in the data sample and expected number of events from
the background sources and the ${\rm N \bar N}$ signal
(Dirac- and Majorana-types)
for the case where ${\rm N \rightarrow e W}$, 
normalized to the actual integrated luminosity at $\protect\sqrt{s}=$~172~GeV.
  The statistical error on the background Monte Carlo is small
  compared to that for data.
}
\label{t:NNe}
\end{center}
\end{table}
\begin{table}
\begin{center}
\begin{tabular}{|c||c|c||c|c|c|c||c|c|} \hline
 After & Data & Total & ${\rm q \bar q}(\gamma)$ & two-photon &$\tau\tau(\gamma)$ &
 4-f   & Dirac & Majorana \\
 Cut   & & Bkg & & & & & $M_{\rm N}=75$ GeV & $M_{\rm N}=65$ GeV \\ \hline
Presel &  $-$ &  $-$ &  $-$ & $-$  & $-$  & $-$   & 6.26 & 4.59  \\
 1     & 1411 & 1288 & 1137 & 26.6 & 3.33 & 120.8 & 5.37 & 3.81  \\
 2     & 0    & 1.23 & 0.27 & 0.0  & 0.0  & 0.96  & 3.41 & 2.53  \\
 3     & 0    & 1.02 & 0.13 & 0.0  & 0.0  & 0.89  & 3.37 & 2.47  \\
 4     & 0    & 0.51 & 0.04 & 0.0  & 0.0  & 0.47  & 3.28 & 2.36  \\ \hline
\end{tabular}
\caption{Observed number of events in the data sample and expected number of events from
  the background sources and the ${\rm N \bar N}$ signal
  (Dirac- and Majorana-types) for the case where
  ${\rm N \rightarrow \mu W}$,
  normalized to the actual integrated luminosity at $\protect\sqrt{s}=$~172~GeV.
  The statistical error on the background Monte Carlo is small
  compared to that for data.
}
\label{t:NNmu}
\end{center}
\end{table}
\begin{table}
\begin{center}
\begin{tabular}{|c||c|c||c|c|c|c||c|c|} \hline
 After & Data & Total & ${\rm q \bar q}(\gamma)$ & two-photon &$\tau\tau(\gamma)$ &
 4-f   & Dirac & Majorana \\
 Cut   & & bkg & & & & & $M_{\rm N}=65$ GeV & $M_{\rm N}=55$ GeV \\ \hline
Presel &  $-$ &  $-$ &  $-$ &  $-$ &  $-$ &  $-$  &10.05 & 8.25 \\
 1     & 1411 & 1288 & 1137 & 26.6 & 3.33 & 120.8 & 9.42 & 7.62 \\
 2     & 4    & 7.99 & 2.44 & 0.59 & 0.10 & 4.84  & 3.68 & 2.86 \\
 3     & 4    & 6.53 & 1.41 & 0.48 & 0.09 & 4.52  & 3.62 & 2.81 \\
 4     & 3    & 3.52 & 0.61 & 0.48 & 0.02 & 2.40  & 3.26 & 2.45 \\
 5     & 2    & 1.98 & 0.21 & 0.0  & 0.01 & 1.76  & 3.05 & 2.26 \\ \hline
\end{tabular}
\caption{Observed number of events in the data sample and expected number of events from
the background sources and the ${\rm N \bar N}$ signal (Dirac and Majorana types) for
the case where $\rm N \rightarrow \tau W$,
normalized to the actual integrated luminosity at $\protect\sqrt{s}=$~172~GeV.
  The statistical error on the background Monte Carlo is small
  compared to that for data.
}
\label{t:NNtau}
\end{center}
\end{table}


\subsubsection{\boldmath Selection of ${\rm L^+L^-}$
  Candidates with ${\rm L^- \rightarrow \nu_\ell W^-}$}
\label{sss:LLnn}

\begin{figure}[b]
\centerline{\hbox{
\epsfig{file=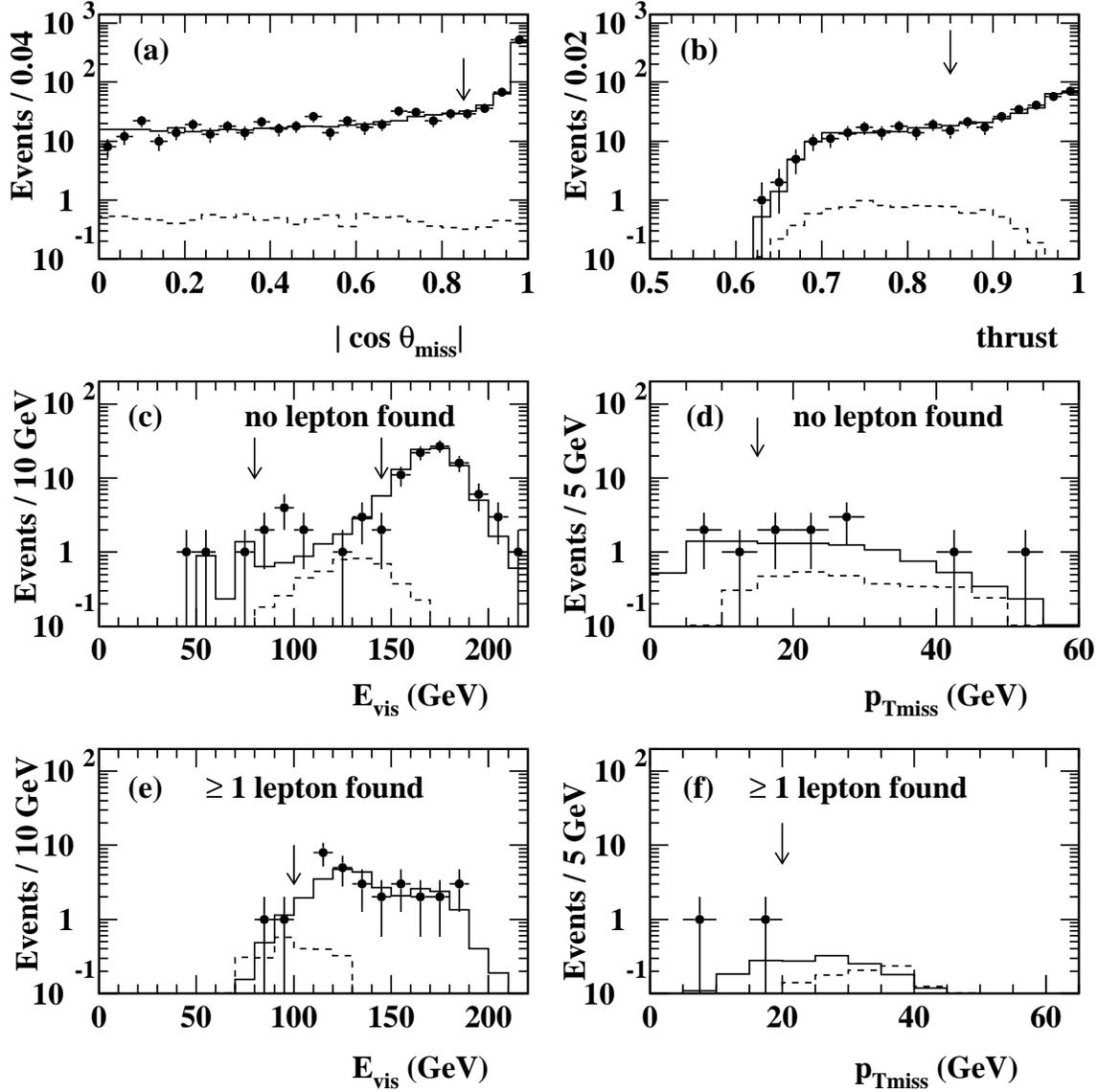,width=\textwidth}
}}
\caption{
  The $|\cos\theta_{\rm miss}|$ distribution after cut~2 (a) and
  the thrust distribution after cut~3 (b),
  for the $\rm L^+L^-$ candidate selection 
  with ${\rm L^-\rightarrow\nu_\ell W^-}$.
  In the case where no isolated lepton is found, the distributions 
  after cut~4 for the visible energy and the missing transverse momentum 
  are shown respectively in (c) and (d).
  In the case where at least one isolated lepton is found,
  the corresponding distributions are shown respectively in (e) and (f).
  The filled circles are the data, the solid lines are the background
  processes, and the broken lines are the $\rm L^+L^-$ signal with
  ${\rm L^-\rightarrow\nu_\ell W^-}$ for the case where $M_{\rm L}=80$ GeV.
  The arrows correspond to the cut values.
}
\label{f:LLnu}
\end{figure}

In order to be considered as potential candidates for ${\rm L^+L^-}$,
or $\ell^{*+}\ell^{*-}$ with CC decays,
where each charged
heavy lepton decays into a light standard neutrino
(category B in Section~\ref{ss:hlintro}, and pair-production
for category B in Section~\ref{ss:elintro}),
the events must satisfy the following set of selection criteria:

\begin{enumerate}
  \item The same multihadronic selection as cut~1 in Section~\ref{sss:NN}.
  \item To reduce the background from ${\rm q \bar q}(\gamma)$ and
        two-photon processes
        we require that the energy deposited in each forward-calorimeter,
        each gamma-catcher,
        and each silicon-tungsten calorimeter be less that 2 GeV, 5 GeV,
        and 5 GeV, respectively.
  \item Non-radiative events are selected  by demanding that
        $|\cos\theta_{\rm miss}| < 0.85$. 
        The $| \cos\theta_{\rm miss} |$ distributions after cut~2 are shown
        in Figure~\ref{f:LLnu}(a).
  \item To reject events containing two back-to-back jets mostly
        from ${\rm q \bar q}(\gamma)$, the thrust of the event
        was required to be less than 0.85.
        The thrust distributions after cut~3 are shown 
        in Figure~\ref{f:LLnu}(b).
  \item If no isolated lepton is found in the event we apply the following constraints:
        \begin{itemize}
          \item $80 < E_{\rm vis} < 145$ GeV.
                The visible energy distributions after cut 4 
                are shown in Figure~\ref{f:LLnu}(c).
          \item $p_{\rm Tmiss} > 15$~GeV, where
                $p_{\rm Tmiss}$ is the
                missing momentum transverse to the beam axis.
                The distributions for $p_{\rm Tmiss}$ after cut~4 are shown in 
                Figure~\ref{f:LLnu}(d).
          \item $N_{\rm jets} \ge 4$, where $N_{\rm jets}$ is the number of reconstructed
                jets.
          \item the maximum momentum of the good tracks should not exceed 25 GeV.
        \end{itemize} 
  \item If one or more isolated lepton is found, we apply the following cuts:
        \begin{itemize}
          \item $E_{\rm vis} < 100$ GeV.
                The visible energy distributions after cut 4 
                are shown in Figure~\ref{f:LLnu}(e).      
          \item $p_{\rm Tmiss} > 20$ GeV.
                The distributions for $p_{\rm Tmiss}$ after cut 4 are shown in 
                 Figure~\ref{f:LLnu}(f).
          \item $N_{\rm jets} \ge 3$.
        \end{itemize}    
\end{enumerate}

The remaining numbers of events after each cut are displayed in
Table~\ref{t:LLB}. The efficiencies obtained after all cuts
are estimated from Monte Carlo to be 20--25\%, depending on the
mass of the heavy or excited lepton.

\begin{table}
\begin{center}
\begin{tabular}{|c||c|c||c|c|c|c||c|} \hline
 After & Data & Total & ${\rm q \bar q}(\gamma)$ & two-photon & $\tau\tau(\gamma)$ &
4-f & ${\rm L^+L^-}$ \\
 Cut      & & Bkg & & & & & $M_{\rm L}=80$ GeV \\  \hline
Presel    &  $-$ &  $-$  &  $-$  &  $-$ &  $-$ &  $-$  & 14.7 \\
    1     & 1411 & 1288  & 1137  & 26.6 & 3.33 & 120.8 & 12.5 \\
    2     & 1044 & 990.3 & 866.5 & 11.2 & 2.84 & 109.9 & 11.5 \\
    3     & 406  & 398.8 & 303.7 & 3.60 & 1.71 &  89.8 & 10.1 \\
    4     & 133  & 129.6 & 70.4  & 1.65 & 0.23 &  57.3 & 7.4  \\
 5 or 6   &   1  & 3.07  &  0.78 & 0.0  & 0.01 &  2.28 & 3.2  \\ \hline
\end{tabular}
\caption{Observed number of events in the data sample and expected number of events from
 the background sources and the ${\rm L^+L^-}$ signal for the case
 where ${\rm L^- \rightarrow \nu_\ell W^-}$,
 normalized to the actual integrated luminosity at $\protect\sqrt{s}=$~172~GeV.
  The statistical error on the background Monte Carlo is small
  compared to that for data.
}
\label{t:LLB}
\end{center}
\end{table}

\subsubsection{\boldmath Selection of ${\rm L^+L^-}$
  Candidates with ${\rm L^- \rightarrow N W^-}$}
\label{sss:llNN}

To be considered as potential ${\rm L^+L^-}$ candidates, where the charged
heavy lepton decays into a stable neutral heavy lepton
(category C in Section~\ref{ss:hlintro}),
the events must satisfy the following set of cuts.  Here we do not
use the standard selection of multihadronic events since for small mass
differences, $M_{\rm L}-M_{\rm N}$, the visible energy is very low and the
particle multiplicity is small.

\begin{figure}[b]
\centerline{\hbox{
\epsfig{file=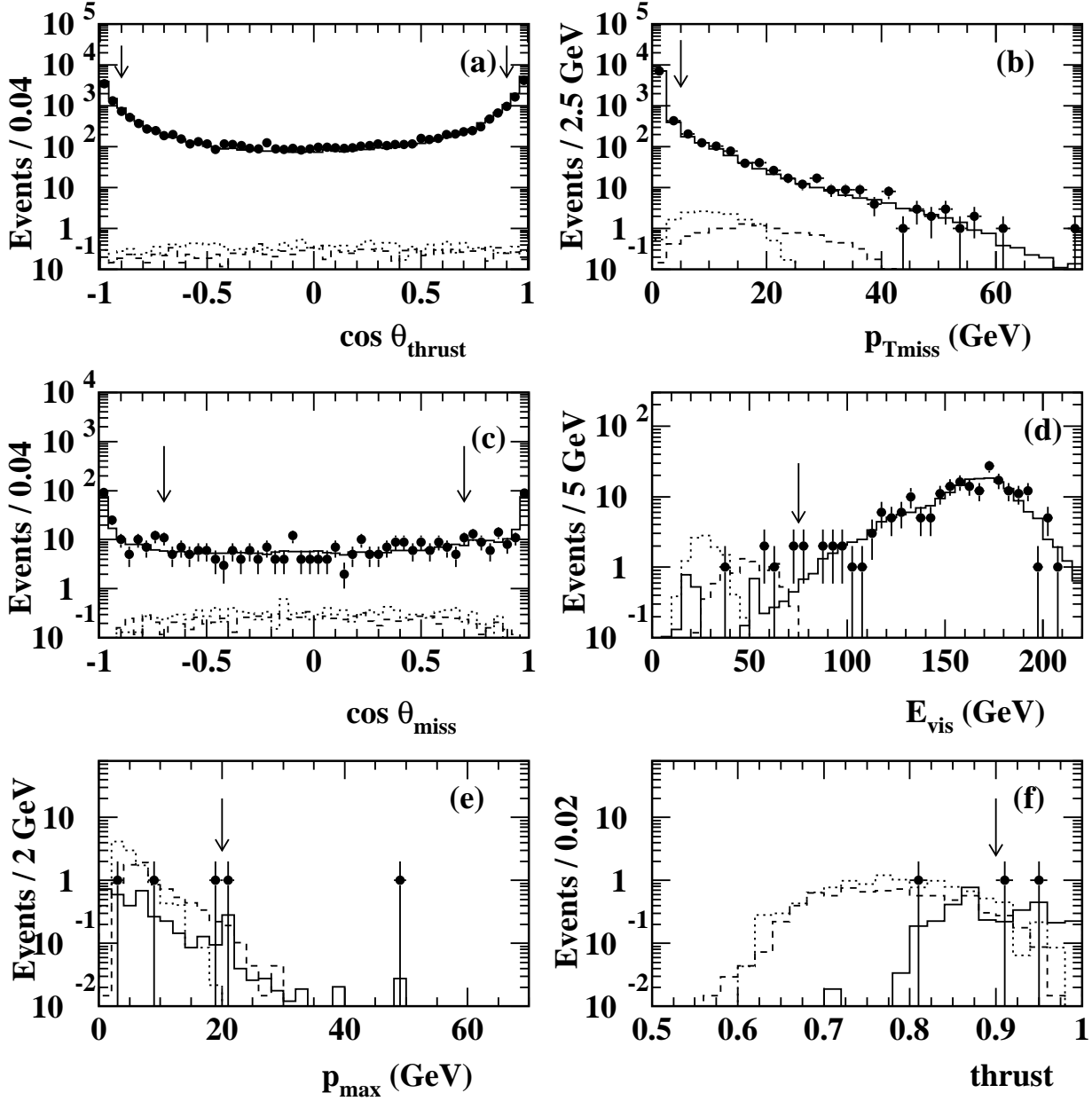,width=\textwidth,
  bbllx=25pt,bblly=150pt,bburx=540pt,bbury=680pt}
}}
\caption{
  The plots show the distributions for the $\rm L^+L^-$ candidate selection
  for the case ${\rm L^-\rightarrow N W^-}$
  for the data (filled circles), the background processes (solid lines), 
  and the $\rm L^+L^-$ signal with ${\rm L^-\rightarrow N W^-}$
  for the cases $(M_{\rm L}=75,M_{\rm N}=60)$ GeV (dotted lines) and
  $(M_{\rm L}=80,M_{\rm N}=50)$ GeV (broken lines $\times 5$)
  respectively for
  $\cos\theta_{\rm thrust}$ after cut~2 (a),
  $p_{\rm Tmiss}$ after cut~3 (b), 
  $\cos \theta_{\rm miss}$ after cut~4 (c),
  $E_{\rm vis}$ after cut~5 (d),
  $p_{\rm max}$ after cut~6 (e),
  and the thrust value after cut~7 (f).
  The arrows correspond to the cut values.
}
\label{f:LLN}
\end{figure}

\begin{enumerate}
  \item We require at least 5 good charged tracks.
  \item The criteria for energy deposited in the forward-detector,
        the gamma-catcher,
        and the silicon-tungsten calorimeter are the same as in 
        Section~\ref{sss:LLnn}.
  \item In order to reduce background from beam-gas and beam-wall interactions
        we require that $|\cos\theta_{\rm thrust}|$ be smaller than 0.9, where
        $\theta_{\rm thrust}$ is the polar angle of the thrust axis.
        The $|\cos\theta_{\rm thrust}|$ distributions after cut 2 are shown in 
        Figure~\ref{f:LLN}(a).
  \item To reduce the large background coming from two-photon processes,
        we apply a cut on the missing momentum transverse to the beam axis:
        $p_{\rm Tmiss} > 5$~GeV (including clusters from
        the hadronic calorimeter), and $p^{\rm no-HCAL}_{\rm Tmiss} > 5$~GeV 
        (excluding  hadronic calorimeter clusters).   
        The $p_{\mathrm Tmiss}$ distributions after cut 3 are
        shown in Figure~\ref{f:LLN}(b).
  \item For radiative-event rejection we require that
        $|\cos\theta_{\rm miss}| < 0.7$.
        The $|\cos\theta_{\rm miss}| $ distributions after cut 4 are shown in 
        Figure~\ref{f:LLN}(c).
  \item The two heavy neutrinos coming from $\rm L^+$ and $\rm L^-$
        carry away a significant fraction
        of the energy. To reduce the background from multihadronic
        and four-fermion events, we require that $ E_{\rm vis}< 75 $ GeV. 
        The $E_{\rm vis}$ distributions after cut 5 are shown in 
        Figure~\ref{f:LLN}(d).
  \item Four-fermion processes and $\tau\tau$ backgrounds were 
        reduced by requiring
        that the maximum momentum of a good track should not exceed 20 GeV.
        The distributions of $p_{\mathrm max}$ after cut 6 are shown
        in Figure~\ref{f:LLN}(e).
  \item We require that the thrust value be less than 0.9 in order 
        to reject events containing two back-to-back jets.
        The distributions of the thrust after cut 7 are shown
        in Figure~\ref{f:LLN}(f).
\end{enumerate}

The remaining numbers of events after each cut are displayed in Table~\ref{t:LLC}. 
The total selection efficiencies are typically 40--50\%, but
drop to 9--12\% for $M_{\rm L}-M_{\rm N}\sim$~5~GeV.
\begin{table}
\begin{center}
\begin{tabular}{|c||c|c||c|c|c|c||c|c|} \hline
 After & Data & Total & ${\rm q \bar q}(\gamma)$ & two-photon &$\tau\tau(\gamma)$ &
 4-f & \multicolumn{2}{|c|}{${\rm L^+L^-}$} \\ \cline{8-9}
Cut & & Bkg & & & & & $M_{\rm L}=75$ GeV & $M_{\rm L}=80$ GeV\\
& &     & & & & & $M_{\rm N}=60$ GeV & $M_{\rm N}=50$ GeV\\ \hline
Presel & $-$ &  $-$  &  $-$  &   $-$ &  $-$ &  $-$  & 21.52 & 14.69 \\
 1     & 40935 & 41602 & 1242  & 40223 & 9.30 & 125.3 & 17.82 & 12.84 \\
 2     & 19792 & 21753 &  908.5& 20725 & 8.02 & 112.1 & 17.30 & 12.34 \\
 3     &  8258 &  8098 &  819.4&  7165 & 7.10 & 106.2 & 15.90 & 11.45 \\
 4     &   531 &  477.1&  382.6&  7.41 & 5.16 &  81.8 & 13.36 & 10.65 \\
 5     &   211 &  200.6& 135.81&  2.04 & 2.60 &  60.2 & 11.15 &  8.44 \\
 6     &    6  &   3.87&   1.10&  1.97 & 0.42 &  0.39 & 11.15 &  8.30 \\
 7     &    3  &   3.35&   1.08&  1.73 & 0.27 &  0.28 & 11.15 &  7.91 \\
 8     &    1  &   1.90&   0.56&  1.12 & 0.06 &  0.16 & 10.33 &  7.35 \\ \hline
\end{tabular}
\caption{Observed number of events in the data sample and expected
  number
  of events from
  the background sources and the ${\rm L^+L^-}$ signal for the case where
  ${\rm L^-\rightarrow N W^-}$, 
  normalized to the actual integrated luminosity at $\protect\sqrt{s}=$~172~GeV.
  The statistical error on the background Monte Carlo is small
  compared to that for data.
  }
\label{t:LLC}
\end{center}
\end{table}


\subsection{Single Production of Excited Leptons with Charged Decays}
\label{ss:spelw}

Two searches are performed for the charged decays of
singly-produced excited leptons
(single-production for categories A and B in Section~\ref{ss:elintro}).
The first analysis looks only for
leptonic decays of the W$^\pm$, which may be real or virtual depending
on the mass of the $\nu^*$.
The topology is then
${\rm e^+e^-} \rightarrow \nu_\ell^*\nu_\ell \rightarrow
 \nu_\ell\ell^\pm{\rm W}^\mp \rightarrow
 \nu_\ell\ell^\pm\nu_{\ell^\prime}\ell^{\prime\mp}$,
where $\ell^\pm$ corresponds to the excited lepton flavour,
and $\ell^{\prime\mp}$ may or may not be of the same flavour.
Since charged excited leptons with charged decays,
${\rm e^+e^-} \rightarrow \ell^{*\pm} \ell^\mp \rightarrow
 \nu_\ell\ell^\mp{\rm W}^\pm \rightarrow
 \nu_\ell\ell^\mp\nu_{\ell^\prime}\ell^{\prime\pm}$,
can have the same final states as neutral excited leptons
with charged decays, the same search is used.
The topology is a pair of leptons which are not coplanar
with the beam direction.

The second search for the charged decays of singly-produced
excited leptons is optimised for the hadronic decays
of the W$^\pm$ boson, looking in the
${\rm e^+e^-} \rightarrow \nu_\ell^*\nu_\ell \rightarrow
 \nu_\ell\ell^\pm{\rm W}^\mp \rightarrow
 \nu_\ell\ell^\pm {\rm q_i \overline{q}_j}$
channels.  In this search, we look only in the electron and
muon channels; the high background levels in the tau channel
do not allow improvement over the sensitivity of
the acoplanar lepton pair analysis.
Since charged excited leptons with charged decays,
${\rm e^+e^-} \rightarrow \ell^{*\pm} \ell^\mp \rightarrow
 \nu_\ell\ell^\mp{\rm W}^\pm \rightarrow
 \nu_\ell\ell^\mp {\rm q_i \overline{q}_j}$,
can have the same final states as neutral excited leptons
with charged decays, the same search is used.

\subsubsection{Acoplanar Lepton Pair Analysis}
\label{sss:spelwll}

\begin{figure}[b]
\centerline{\hbox{
\epsfig{file=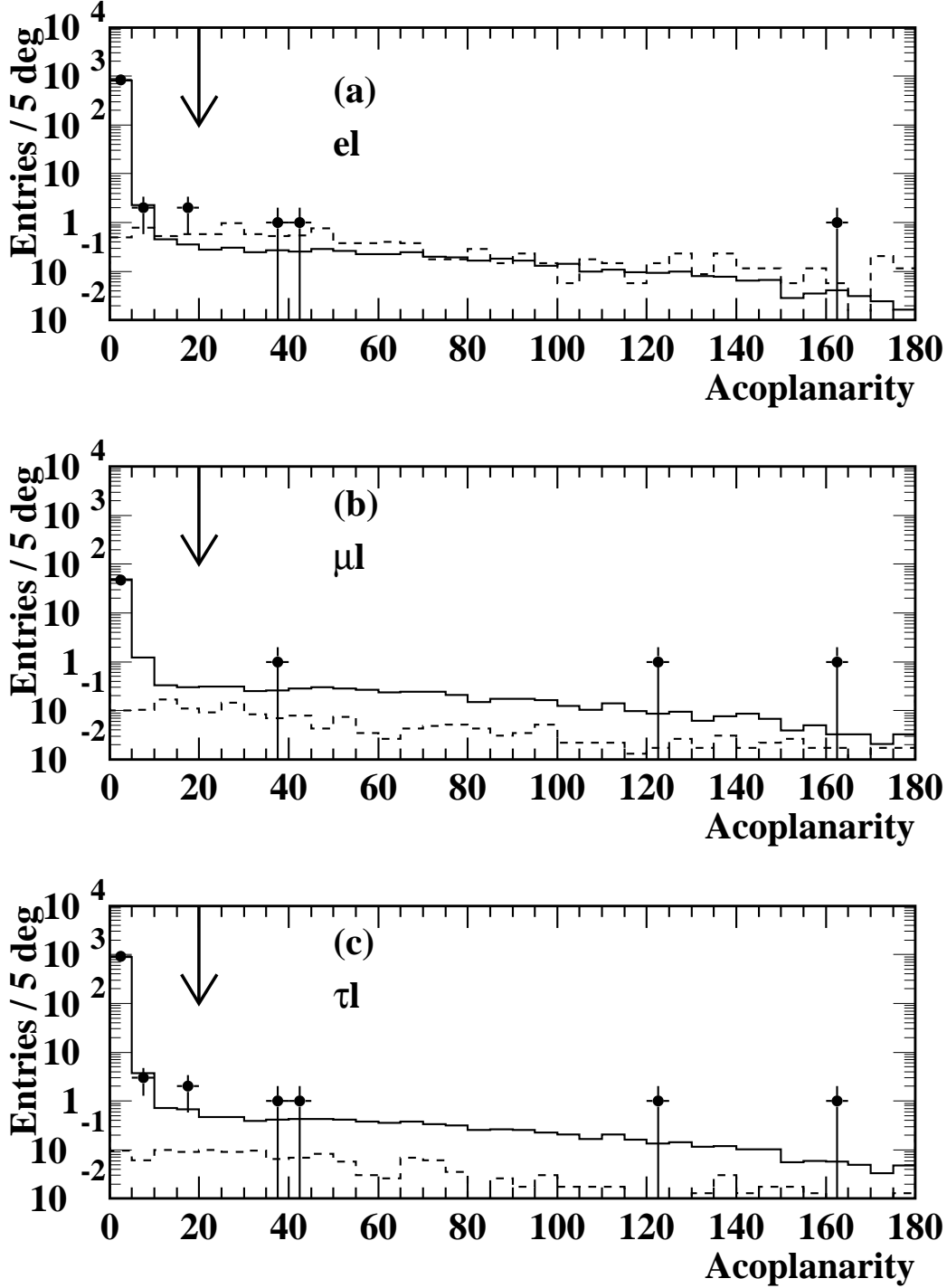,width=14cm,
  bbllx=100pt,bblly=150pt,bburx=470pt,bbury=680pt}
}}
\caption{
  $\ell^*$ single production, charged decays:
  the acoplanarity angle between the two observed
  leptons, before the cut on acoplanarity angle.
  (a) is for ${\rm e}^*$ and $\nu_{\rm e}^*$,
  (b) is for $\mu^*$ and $\nu_\mu^*$ and
  (c) is for $\tau^*$ and $\nu_\tau^*$.
  The solid line is the sum of all Standard Model
  background Monte Carlo,
  the dashed lines are $\ell^{*\pm}\ell^\mp$ (charged decay)
  Monte Carlo with a mass of 80 GeV
  normalized to $f/\Lambda=(200~{\rm GeV})^{-1}$,
  and the filled circles are the data.
  The arrows show the cut value of 20$^\circ$.
  }
\label{f:acopll}
\end{figure}

Using the same preselection and track quality cuts
as the $\ell^+\ell^-\gamma(\gamma)$ analysis
in Section~\ref{ss:celg},
candidate $\ell^{*\pm}$ and $\nu_\ell^*$ events are required to satisfy the
following criteria:
\begin{enumerate}
  \item There must be exactly two identified leptons, at least one of type
        $\ell^\pm$ corresponding to the excited lepton flavour.
        At least one of the leptons must have an energy
        greater than $0.2E_{\rm beam}$.
  \item There must be no other track passing the quality cuts in the event.
  \item Excluding the clusters associated with the two leptons,
        the sum of other barrel and endcap
        electromagnetic calorimeter cluster energies must
        be less than 10~GeV.
  \item Events from two-photon processes
        are suppressed by requiring that the energy deposited in each
        forward-calorimeter, each gamma-catcher and each silicon-tungsten
        calorimeter be less than 2~GeV, 5~GeV and 5~GeV, respectively.
  \item Events with missing momentum along the beam axis are removed by
        requiring $|\cos{\theta_{\rm miss}}|<0.9$, where
        $\theta_{\rm miss}$ is the polar angle of the
        missing momentum.
  \item Events are required to be acoplanar by requiring
        $\phi^{\rm ACOP}_{\ell\ell}>20^\circ$
        ($\phi^{\rm ACOP}_{\ell\ell}$ is 180$^\circ$ minus the opening
        angle between the two leptons in the $x$-$y$ plane).
        The acoplanarity angle is plotted for the excited electron,
        muon and tau search in Figures~\ref{f:acopll}(a), (b) and (c),
        respectively.
\end{enumerate}

The complete analysis is summarized in Table~\ref{t:lnuwll}.
The numbers of events selected in each channel are
consistent with the expectations from Standard Model sources.
After cut~6, the dominant background is from ${\rm W^+W^-}$
production.
The total selection efficiencies for excited lepton single
production with CC decays when the W$^\pm$ decays leptonically
is about 55\% for ${\rm e^*}$, $\nu_{\rm e}^*$,
$\mu^*$ and $\nu_\mu^*$, and about 45\% for
$\tau^*$ and $\nu_\tau^*$.  

\begin{table}
\begin{center}
\begin{tabular}{|c|c|r|r||c|c|c|c|} \hline
  Topology            & After &  Data  & Total    & \multicolumn{2}{c|}{$\ell^{*\pm}\ell^\mp$} & \multicolumn{2}{c|}{$\nu^*\nu$}   \\ \cline{5-8}
                        & Cut &        &  Bkg     &    $M_*=$        &      $M_*=$             &     $M_*=$          &   $M_*=$    \\
                        &     &        &          &    80 GeV        &      140 GeV            &     80 GeV          &   140 GeV   \\ \hline
                        &  1  &  1377  &  1344.3  &      23.7        &      23.4               &      20.5           &    23.8     \\
                        &  2  &  1356  &  1330.8  &      23.3        &      23.3               &      20.1           &    23.6     \\
 ${\rm e}\ell$          &  3  &  1297  &  1272.7  &      22.0        &      23.0               &      19.2           &    23.4     \\
                        &  4  &  1208  &  1202.1  &      20.3        &      22.3               &      18.7           &    23.2     \\
                        &  5  &   852  &   833.0  &      18.9        &      20.4               &      17.9           &    21.4     \\
                        &  6  &     3  &     4.7  &      14.8        &      18.2               &      16.9           &    19.8     \\ \hline
                        &  1  &   192  &   189.8  &      25.7        &      26.5               &      22.3           &    25.3     \\
                        &  2  &   185  &   184.1  &      25.3        &      26.3               &      22.1           &    25.2     \\
 $\mu\ell$              &  3  &   163  &   156.0  &      23.5        &      26.0               &      20.6           &    24.7     \\
                        &  4  &   136  &   129.0  &      22.1        &      25.5               &      18.9           &    24.3     \\
                        &  5  &    50  &    55.7  &      20.0        &      24.2               &      17.9           &    22.0     \\
                        &  6  &     3  &     5.0  &      14.4        &      21.6               &      17.0           &    19.9     \\ \hline
                        &  1  &  1646  &  1621.6  &      21.2        &      18.8               &      14.8           &    22.9     \\
                        &  2  &  1623  &  1603.5  &      20.9        &      18.7               &      14.6           &    22.9     \\
 $\tau\ell$             &  3  &  1536  &  1513.7  &      19.6        &      18.4               &      13.5           &    22.6     \\
                        &  4  &  1415  &  1407.8  &      18.3        &      18.0               &      12.5           &    22.2     \\
                        &  5  &   930  &   913.6  &      16.5        &      16.7               &      12.0           &    20.8     \\
                        &  6  &     4  &     7.5  &      12.5        &      14.7               &      11.2           &    19.1     \\ \hline
\end{tabular}
\end{center}
\caption{
  Number of events surviving each cut, and selection
  efficiencies in percent, in the
  search for the single production of 
  $\ell^{*\pm}$ and $\nu^*$ with charged decays in
  the acoplanar di-lepton channel.
  The background Monte Carlo sum is
  normalized to the actual integrated luminosity at $\protect\sqrt{s}=$~172~GeV.
  The three selections are not independent; in particular, the
  $\tau\ell$ selection completely includes both the
  ${\rm e}\ell$ and $\mu\ell$ selections.
  The efficiencies include the leptonic branching
  ratio of the W.
  The statistical error on the background Monte Carlo is small
  compared to that for data.
  }
\label{t:lnuwll}
\end{table}

\subsubsection{\boldmath $\ell^\pm \nu {\rm q \bar q}$ Analysis}
\label{sss:spelwljj}

\begin{figure}[b]
\centerline{\hbox{
\epsfig{file=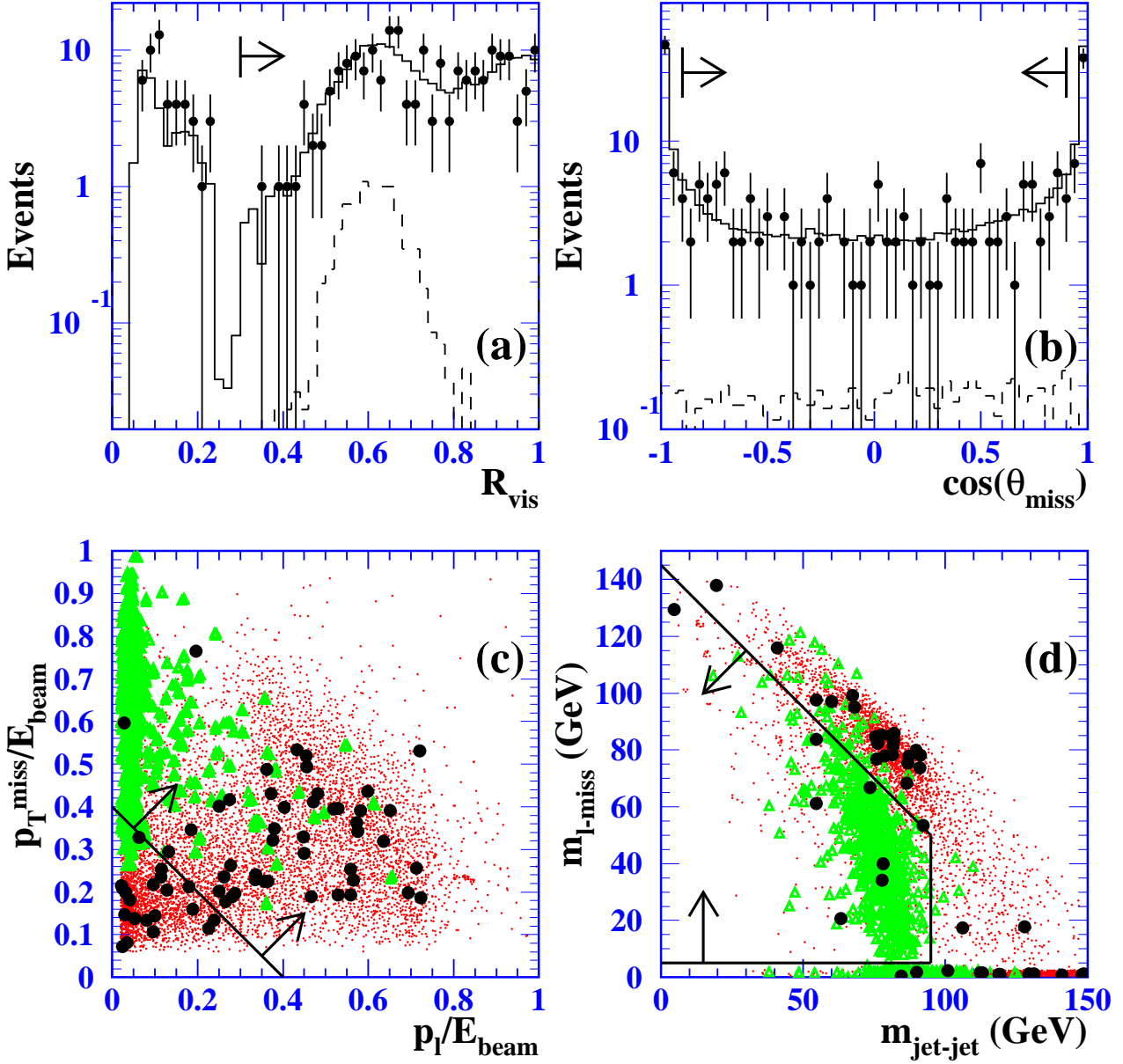,width=\textwidth,
  bbllx=25pt,bblly=150pt,bburx=540pt,bbury=680pt}
}}
\caption{
  Single-production of charged excited leptons with CC decays,
  with the cuts applied sequentially.
  (a) is the event visible energy scaled to $\protect\sqrt{s}$, and
  (b) is cosine of the polar angle of the missing momentum.
  In (a) and (b), the filled circles are the data, the solid line is
  the sum of all Standard Model background Monte Carlo, and the
  dashed line is excited electron single-production Monte
  Carlo with $M_*=$~170~GeV and arbitrary normalisation.
  (c) is the component of the missing momentum transverse to
  the beam axis {\it vs.} the lepton momentum, and
  (d) is the invariant mass of the lepton-neutrino {\it vs.} the
  invariant mass of the jet-jet pair.
  In (c) and (d), the filled circles are the data, the dots
  are the Standard Model background Monte Carlo, and the 
  triangles are excited electron single production Monte
  Carlo with $M_*=$~170~GeV.
  The arrows indicate the region accepted by the cuts on
  these quantities.
}
\label{f:lnuqq}
\end{figure}

All charged tracks and calorimeter clusters are subjected to the same quality
criteria used in the unstable heavy lepton analysis in Section~\ref{ss:pphl}.
As in that section, overlap between tracks and clusters is also corrected 
using the method described in Reference~\cite{ref:gce}.
A preselection is used to remove low multiplicity events.
At least six tracks passing the quality cuts must be reconstructed in an event,
the ratio of the number of selected tracks to the total number of 
reconstructed tracks must be 
greater than 0.2, and at least 8 clusters must be reconstructed in the event.
Following the preselection, candidate events are required to satisfy
the following criteria:

\begin{enumerate}
  \item There must be at least one isolated lepton $\ell$
        ($\ell = {\mathrm e}$ or $\mu$), corresponding to
        the excited lepton flavour.  
        The reconstructed track corresponding to the lepton candidate
        should lie within the angular region $|\cos{\theta}| < 0.95$.
        Leptons consistent with coming from a Z decay
        are rejected if their invariant mass when combined with
        any other track
        is within 10 GeV of the Z mass. 
        If there is more than one lepton candidate, then the
        candidate with the highest momentum is taken.
        The lepton must satisfy one of the following criteria:
        \begin{description}
          \item[electron:] A track is identified as an electron if it satisfies
                           either of the following criteria:
            \begin{enumerate}
              \item The track must be associated with an electromagnetic calorimeter cluster
                    with a minimum energy of 2.0 GeV.
                    The sum of the magnitudes of the momenta of tracks and energy of
                    unassociated electromagnetic clusters in a cone with half-angle
                    of 15$^{\circ}$ around the electron track should be less than 2.5 GeV.
                    Candidate electron tracks with
                    $|\cos{\theta}| < 0.79$ or  $|\cos{\theta}| \geq 0.815$,
                    are required to satisfy $E/p > 0.7$.  Candidate electron tracks with
                    $0.79 \leq |\cos{\theta}| < 0.815$ should satisfy $E/p > 0.5$.
                    The ${\rm d}E/{\rm d}x$ energy loss of tracks must be consistent
                    with that expected from an electron.
              \item The output of a neural network electron finder~\cite{ref:NN} must be greater than 0.8.
            \end{enumerate}
           \item[muon:] A track with $p>2$~GeV is identified as a muon if it is matched                        
                        with track segments in the muon chambers.
                        In regions not covered
                        by the muon chambers the identification is done by using the
                        hadron calorimeter strips.
        \end{description}
      \item To remove events from two-photon processes, events having significant
            energy in either side of the forward detectors are 
            rejected.  The maximum energy allowed in the gamma catcher is 5 GeV,
            in the forward calorimeter 2 GeV, and in the silicon tungsten detector is 2 GeV. 
      \item To reduce further two-photon events, the ratio of total visible energy 
            of the event to the centre-of-mass energy, defined as 
            $R_{\rm vis} = E_{\rm vis}/\sqrt{s}$, 
            is required to be greater than 0.3.  $R_{\rm vis}$ is plotted before this cut
            in Figure~\ref{f:lnuqq}(a).
      \item To reject events with initial state radiation along the beam axis, 
            the missing energy vector of the event
            must satisfy $|\cos{\theta_{\rm miss}}|<0.9$;
            $\cos{\theta_{\rm miss}}$ is plotted before the cut in Figure~\ref{f:lnuqq}(b).
      \item To suppress further the background from hadronic Z decays, 
            the thrust of the event must be less than 0.95. 
      \item The event is required to have a large missing momentum, $p_{\rm Tmiss}$, or a large lepton
            energy, $p_{\ell}$, with respect to the beam energy, $E_{\rm beam}$.  This is achieved by
            requiring $( p_{\rm Tmiss} + p_{\ell} ) / E_{\rm beam} > 0.4$.
            $p_{\rm Tmiss}$ {\it vs.} $p_{\ell}$ is plotted before this cut in Figure~\ref{f:lnuqq}(c).
            While only one example mass for an $\ell^*$ signal is shown in the figure, this
            cut retains high efficiency for all $\ell^*$ masses.
      \item To reduce the contribution from W-pair production,
            the lepton candidate is removed from the event and the event
            is forced into two jets, using the Durham jet algorithm.
            The resulting invariant masses $m_{\ell - {\rm miss}}$ and $m_{\rm {jet-jet}}$
            are required to satisfy:
            \begin{itemize}
              \item $m_{\ell - {\mathrm miss}} > 5$~GeV 
                    and $m_{\mathrm jet-jet} < 95$~GeV,
              \item $m_{\ell - {\mathrm miss}} 
                    + m_{\mathrm jet-jet} < 145$~GeV.
            \end{itemize}
            $m_{\ell - {\rm miss}}$ {\it vs.} $m_{\rm {jet-jet}}$ is plotted before this cut
            in Figure~\ref{f:lnuqq}(d).
\end{enumerate}

The complete analysis is summarized in Table~\ref{t:lnuwqq}.
The numbers of events selected in each channel are
consistent with the expectation from Standard Model sources.
The total selection efficiency for excited lepton single
production with CC decays when the W$^\pm$ decays hadronically
is typically from 20--45\%, depending on the mass of the excited lepton,
but drops to about 10\% for $\nu^*$ near the kinematic
limit.

\begin{table}
\begin{center}
\begin{tabular}{|c|c|r|r||c|c|c|c|} \hline
  Topology            & After &  Data  & Total    & \multicolumn{2}{c|}{$\ell^{*\pm}\ell^\mp$} & \multicolumn{2}{c|}{$\nu^*\nu$} \\ \cline{5-8}
                        & Cut &        &  Bkg     &    $M_*=$        &      $M_*=$             &     $M_*=$          &   $M_*=$  \\
                        &     &        &          &    80 GeV        &      140 GeV            &     80 GeV          &   140 GeV \\ \hline
                        &  1  &   449  &   403.3  &    59.4          &       64.8              &     59.7            &    61.2    \\
                        &  2  &   271  &   264.2  &    52.5          &       60.0              &     55.5            &    57.0    \\
 ${\rm e}{\rm q\bar q}$ &  3  &   223  &   233.6  &    52.1          &       60.0              &     55.5            &    57.0    \\
                        &  4  &   119  &   119.7  &    42.8          &       54.3              &     52.0            &    53.7    \\
                        &  5  &    98  &    96.5  &    39.6          &       54.0              &     51.5            &    51.9    \\
                        &  6  &    45  &    38.0  &    31.8          &       53.7              &     51.2            &    50.8    \\
                        &  7  &     7  &     4.4  &    15.2          &       23.2              &     30.8            &    22.0    \\ \hline
                        &  1  &   103  &   114.6  &    51.8          &       52.4              &     49.5            &    52.2    \\
                        &  2  &    45  &    57.8  &    45.8          &       48.2              &     44.3            &    48.6    \\
 $\mu{\rm q\bar q}$     &  3  &    30  &    36.2  &    45.6          &       48.2              &     44.2            &    48.6    \\
                        &  4  &    20  &    23.2  &    35.9          &       43.8              &     37.1            &    44.6    \\
                        &  5  &    18  &    21.6  &    34.6          &       43.6              &     36.7            &    43.6    \\
                        &  6  &     9  &    12.9  &    27.6          &       43.4              &     35.8            &    43.2    \\
                        &  7  &     1  &     0.6  &    14.4          &       18.4              &     20.9            &    18.0    \\ \hline
\end{tabular}
\end{center}
\caption{
  Number of events surviving after each cut, and the
  selection efficiencies in percent, in the
  search for the single-production of 
  $\ell^{*\pm}$ and $\nu^*$ with CC decays in
  the lepton plus jets and missing energy analysis.
  The background Monte Carlo sum is
  normalized to the actual integrated luminosity at $\protect\sqrt{s}=$~172~GeV.
  The efficiency includes the hadronic branching
  ratio of the W.
  The statistical error on the background Monte Carlo is small
  compared to that for data.
  }
\label{t:lnuwqq}
\end{table}

No signal is observed.
The results from both the acoplanar lepton pair and 
$\ell^\pm\nu{\rm q\bar q}$ analyses are combined to infer 
limits on excited lepton single production with CC decays.
The $\ell^\pm\nu{\rm q\bar q}$ results include
a $\pm$~3~$\sigma$ mass window, where $\sigma$ is the average
mass resolution determined from Monte Carlo, to classify each event
as being consistent with an excited lepton in some mass range,
while the selected acoplanar lepton pair events
are considered as candidates for all excited 
lepton masses.


\subsection{Production of Neutral Excited Leptons with Photonic Decays}
\label{ss:nelg}

The search for excited neutrino production with photonic decays 
(category C in Section~\ref{ss:elintro})
uses the OPAL analysis of photonic events with missing
energy \cite{ref:ggOPAL172}.
That analysis has one method optimised for events with one photon
plus missing energy and another method optimised for
events with two photons plus missing energy.
These are used for the excited neutrino single- and 
pair-production searches, respectively.
The search results for the full $\sqrt{s}=$~161--172~GeV
data set are summarized in Table~\ref{t:ggsearch}; these selections
have improved sensitivity over our previously published 
analyses~\cite{ref:elopal161}, and are therefore used to update
our $\sqrt{s}=$~161~GeV results as well.

The search for a single photon plus missing energy
selects events with either a single, energetic isolated
cluster in the electromagnetic calorimeter, or
an identified photon conversion, along with
no other significant energy in the event.
One event is selected at  $\sqrt{s}=$~161~GeV
and another at $\sqrt{s}=$~172~GeV,
with a combined expected background of 0.7 to 0.8~events from
Standard Model processes.  There is a discrepancy of about
15\% between the predictions of the expected background level
to this analysis from the KORALZ and NUNUGPV generators,
and we use the smaller number when performing limit calculations.
The search for a photon pair plus missing energy selects
events with two photon candidates, along with no other
significant energy in the event.
One event is selected at  $\sqrt{s}=$~161~GeV, and
two more events are selected at $\sqrt{s}=$~172~GeV,
with an expected background from 4.7 to 9.2~events
from Standard Model processes.
In the photon pair topology, the discrepancy between
KORALZ and NUNUGPV is much more serious, and no
background subtraction is performed.
As described in Reference~\cite{ref:ggOPAL172}, the
kinematics of the photons are used to classify
the events as being consistent with excited neutrino
pair-production up to a maximum mass, $M^{\rm max}$. The 
maximum excited neutrino mass consistent with any of the 
three events is 69.7~GeV.  
While the events are taken as candidates, 
the excited neutrino masses with which they are
consistent are sufficiently low
that they do not affect the limits inferred from
these data.

The total selection efficiency for photonic decays of
pair-produced excited neutrinos is about 80\%,
and for singly-produced excited neutrinos varies
from about 20--80\% for excited neutrino masses
of 80--170~GeV.  The selection efficiency for
singly-produced excited electron-neutrinos is
slightly different than for excited muon- and tau-neutrinos 
because of the $t$-channel W$^\pm$-exchange
contribution to the production angular distribution.

\begin{table}
\begin{center}
\begin{tabular}{|c|c|c||c|c|} \hline
 Topology   &        Requirements            & Data  &  NUNUGPV  & KORALZ             \\ \hline
            & Single $\gamma$ selection      &  85   &    94.7   &  92.6              \\ \cline{2-5}
 1 $\gamma$ & $M_{\rm miss}<$~75 GeV         &   3   &     1.7   &   1.9              \\ \cline{2-5}
            & Remove poorly measured regions &   2   &     0.8   &   0.7              \\ \hline\hline
 2 $\gamma$ & Two $\gamma$ selection         &   3   &     9.2   &   4.7              \\ \hline
\end{tabular}
\end{center}
\caption{
  Number of events surviving in the two photonic events topologies
  at $\protect\sqrt{s}=$~161--172~GeV.
  The Standard Model background is completely dominated by
  $\nu\bar\nu\gamma\gamma$, and predictions from both the
  NUNUGPV and KORALZ generators are shown.
  The analyses and requirements used in new particle searches
  are discussed in detail in Reference~\cite{ref:ggOPAL172}.
  The statistical error on the background Monte Carlo is small
  compared to that for data.
  }
\label{t:ggsearch}
\end{table}


\subsection{Production of Charged Excited Leptons with Photonic Decays}
\label{ss:celg}

The single- and pair-production of excited leptons with photonic
decays leads to $\ell^+\ell^-\gamma(\gamma)$ topologies
(category D in Section~\ref{ss:elintro}).
To be considered in the analysis, tracks in the central
detector and clusters in the  electromagnetic calorimeter 
must satisfy the normal quality criteria employed in the
analysis of lepton pairs~\cite{ref:LL}.
In addition, a ``good'' track must satisfy $|\cos{\theta}|<0.95$.

A pre-selection is performed to remove obvious background events.
Cosmic rays are rejected
using timing and tracking information as described
in the analysis of lepton pairs \cite{ref:LL}. 
Residual beam-gas and beam-wall collision events are rejected
by requiring that the fraction of good to total charged tracks
reconstructed in the central detector be greater than 0.2.
Background from multihadronic events is reduced by requiring that
the number of good tracks (after removing tracks
identified as part of photon conversions as described below)
satisfies $N_{\rm trk}\leq$~4.
The pre-selection also requires a minimum of one lepton candidate
(electron, muon, or tau) and a minimum of
one photon candidate, in the event,
using the criteria defined below.

Tracks with $p_{\rm T}>$~1~GeV are considered as potential
lepton candidates, and the following identification requirements
are made:
\begin{description}
  \item[electron:] A track is identified as an electron if it satisfies
    any one of the following three criteria:
    \begin{enumerate}
      \item $0.8<E/p<1.3$, where $p$ is the momentum of the track and $E$
            is the energy of the associated electromagnetic cluster.
      \item $0.5<E/p<2.0$ and the signed ${\rm d}E/{\rm d}x$
            weight~\cite{ref:dedx} is consistent with the track
            being an electron.
      \item The output of the electron identification neural network
            described in 
            Reference~\cite{ref:NN} is greater than 0.8.
            This final criterion is desirable to recover some extra
            efficiency for low-energy recoil electrons in the 
            single-production search.
    \end{enumerate}
    The energy, $E_{\rm e}$, of the electron is computed using
    the electromagnetic calorimeter cluster energy and its direction
    is computed using the track observed in the central detector.
  \item[muon:] A track is identified as a muon if it satisfies
               either of the following two criteria:
    \begin{enumerate}
      \item The track is identified as a muon according to the
            criteria employed in the
            analysis of Standard Model muon pairs~\cite{ref:LL};
            that is, it
            has associated activity in the muon chambers or hadron
            calorimeter strips or it has a high momentum but is
            associated with only a small energy deposit
            in the electromagnetic calorimeter.
      \item The track is identified as a muon according to the
            criteria employed in the analysis of muons in
            multihadronic events given in Reference~\cite{ref:MU}.
            This second criterion is desirable to recover some extra
            efficiency for low-energy recoil muons in the 
            single-production search.
    \end{enumerate}
  \item[tau:] a track is identified as originating from a tau decay
              if it satisfies either of the following two conditions:
    \begin{enumerate}
      \item It is identified as an electron or muon according to the
            above requirements (i.e. electrons and muons are also
            used as tau candidates).
      \item There are at most two additional tracks
            in a cone with a half-angle 
            of $20^\circ$ around the track.
    \end{enumerate}
    A tau ``jet'' is constructed by adding the momenta of all tracks and
    clusters inside the $20^\circ$ half-angle cone.  The energy of
    the tau jet is calculated using only the tracks in the
    central detector and energy in the electromagnetic calorimeter,
    correcting for double-counting using the method described
    in Reference~\cite{ref:gce}.
\end{description}

A separate search is performed to identify photons:
\begin{description}
  \item [photon:] A photon candidate must satisfy either of the following
        two criteria:
    \begin{enumerate}
      \item An electromagnetic cluster with no associated good charged tracks.
      \item A photon conversion identified with the algorithm
            employed in the analysis of
            muon pairs \cite{ref:LL}.  The tracks and clusters
            associated with the conversion are combined to form a single
            4-vector representing the photon.
    \end{enumerate}
    The photon is also required to satisfy
    $|\cos{\theta}|<0.95$ and to have an energy greater than 1~GeV.
\end{description}

\subsubsection{\boldmath $\ell\ell\gamma\gamma$ Final States }
\label{sss:llgg}

\begin{figure}[b]
\centerline{\hbox{
\epsfig{file=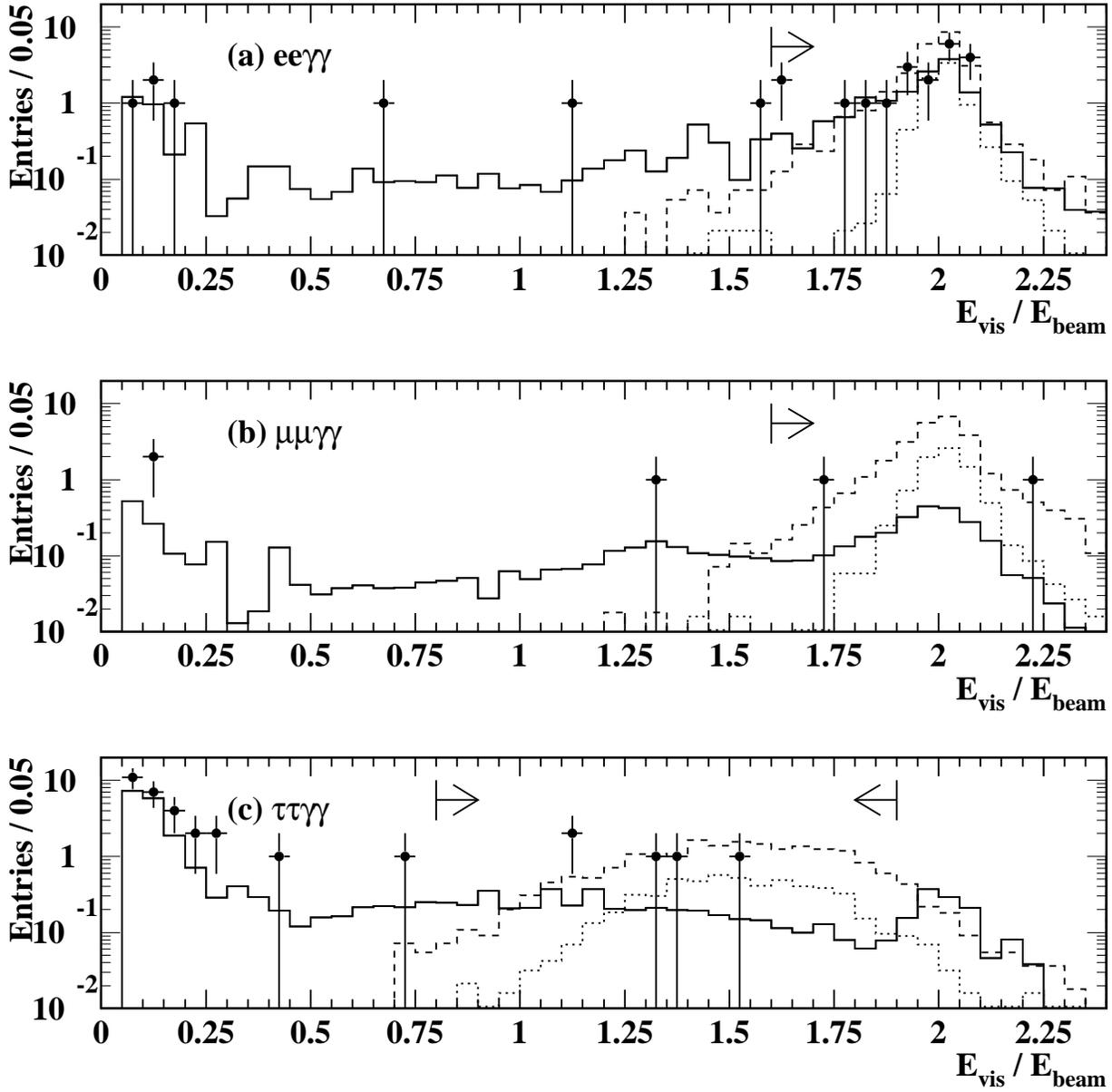,width=\textwidth,
  bbllx=25pt,bblly=150pt,bburx=540pt,bbury=680pt}
}}
\caption{
  The sums of the energies of the two most
  energetic leptons and two most energetic photons,
  after requiring at least two leptons of the same
  flavour and two photons in the event.
  (a) is for ${\rm e^+e^-}\gamma\gamma$,
  (b) is for $\mu^+\mu^-\gamma\gamma$    and
  (c) is for $\tau^+\tau^-\gamma\gamma$.
  The dashed and dotted lines are
  $\ell^{*+}\ell^{*-}$ photonic-decay
  signal Monte Carlo
  with masses of 70 and 85~GeV, respectively,
  the solid lines are the sum of all of the Standard Model background
  Monte Carlo and the filled
  circles are the data.
  The arrows indicate the cut positions.
}
\label{f:llggen}
\end{figure}

Candidate $\ell^+\ell^-\gamma\gamma$ events are required to satisfy the
following criteria:

\begin{enumerate}
  \item There must be at least two identified leptons of the same
        flavour and at least two photons.
        The two most energetic leptons and
        two most energetic photons are used for further analysis.
        For the excited tau search, it is required that at most one
        of the two tau jets be an identified electron, and at most one
        of the two tau jets be an identified muon.
  \item In the ${\rm e^+e^-}\gamma\gamma$ and $\mu^+\mu^-\gamma\gamma$
        analyses, the sum of the energies of the two leptons and two photons,
        $E_{\rm vis}$, must satisfy
        $E_{\rm vis} > 1.6~E_{\rm beam}$.
        In the $\tau^+\tau^-\gamma\gamma$
        analysis, the sum of the energies of the two tau jets
        and two photons must satisfy
        $0.8~E_{\rm beam} < E_{\rm vis} < 1.9~E_{\rm beam}$.
        The sums of the energies of the two leptons
        and two photons before cut~2 are plotted in Figure~\ref{f:llggen}.
  \item A significant background after cuts~1 and 2 is from Bhabha scattering
        or lepton-pair production with final-state radiation.
        This is reduced by requiring the lepton-photon pair to
        be isolated.  The minimum opening angle among all
        lepton-photon combinations, $\theta_{\rm min}^{\ell\gamma}$,
        is required to satisfy
        $|\cos\theta_{\rm min}^{\ell\gamma}|<$~0.90 for
        ${\rm e^+e^-}\gamma\gamma$ and 
        $|\cos\theta_{\rm min}^{\ell\gamma}|<$~0.95 for $\mu^+\mu^-\gamma\gamma$
        and $\tau^+\tau^-\gamma\gamma$.
  \item The momenta of the two leptons and two photons are now
        calculated assuming a 4-body final state with no
        missing energy, using the measured angles of the tracks
        and clusters.  The calculation uses the beam-energy
        constraint, conserving energy and momentum.
        It is required that all energies calculated are greater
        than zero.  The momenta of the leptons calculated in
        this manner are used in cut~5.
  \item The final reducible component in the background
        is from the doubly-radiative-return process
        ${\rm e^+e^-}\rightarrow {\rm Z^0}\gamma\gamma
        \rightarrow \ell^+\ell^-\gamma\gamma$.
        Events with a di-lepton mass,
        $M_{\ell\ell}$, satisfying
        80~GeV~$<M_{\ell\ell}<$~100~GeV are vetoed.
\end{enumerate}

The complete analysis is summarized in Table~\ref{t:psum},
including an example signal of charged excited leptons
with photonic decays.
No event survives in any channel,
which is consistent with the expectations from Standard
Model sources.

The efficiency for observing the pair-production of excited
leptons with photonic decays is estimated with Monte Carlo
to be about 54\% for ${\rm e^{*+}e^{*-}}$, 62\% for $\mu^{*+}\mu^{*-}$
and 42\% for $\tau^{*+}\tau^{*-}$, with only a small dependence
on the mass of the excited lepton.

\begin{table}
\begin{center}
\begin{tabular}{|c|c||c|c||c|c|c||c|} \hline
                        & After & Data & Total  &  ee$(\gamma)$   & $\mu\mu(\gamma)$ & $\tau\tau(\gamma)$ & $M_* =$ \\
                        & Cut   &      &  Bkg   &       &          &            & 80 GeV \\ \hline
                        &  1  &   33 &   23.9 &  18.8 &     0.0  &    1.5     &  69.9    \\
                        &  2  &   26 &   17.2 &  17.0 &     0.0  &    0.1     &  69.6    \\
 ${\rm ee}\gamma\gamma$ &  3  &    3 &    2.3 &   2.2 &     0.0  &    0.0     &  60.7    \\
                        &  4  &    2 &    1.5 &   1.5 &     0.0  &    0.0     &  58.4    \\
                        &  5  &    0 &    1.3 &   1.3 &     0.0  &    0.0     &  53.0    \\ \hline
                        &  1  &    5 &    5.5 &   0.0 &     3.4  &    0.8     &  75.5    \\
                        &  2  &    2 &    2.6 &   0.0 &     2.5  &    0.1     &  75.2    \\
 $\mu\mu\gamma\gamma$   &  3  &    0 &    1.4 &   0.0 &     1.4  &    0.0     &  70.2    \\
                        &  4  &    0 &    1.1 &   0.0 &     1.1  &    0.0     &  67.7    \\
                        &  5  &    0 &    0.8 &   0.0 &     0.8  &    0.0     &  61.3    \\ \hline
                        &  1  &   33 &   23.8 &   1.6 &     0.1  &    5.0     &  54.3    \\
                        &  2  &    5 &    4.2 &   0.3 &     0.0  &    3.4     &  51.4    \\
 $\tau\tau\gamma\gamma$ &  3  &    1 &    1.2 &   0.0 &     0.0  &    0.9     &  49.1    \\
                        &  4  &    0 &    0.7 &   0.0 &     0.0  &    0.6     &  45.7    \\
                        &  5  &    0 &    0.5 &   0.0 &     0.0  &    0.5     &  41.8    \\ \hline
\end{tabular}
\end{center}
\caption{
  The number of events surviving each cut for background, and
  selection efficiencies in percent for
  some example signal Monte Carlos, in the $\ell^+\ell^-\gamma\gamma$
  topology.  The efficiencies are for the pair production
  of charged excited leptons which decay photonically.
  The expected background levels are
  normalized to the actual integrated luminosity.
  The statistical error on the background Monte Carlo is small
  compared to that for data.
  }
\label{t:psum}
\end{table}

\subsubsection{\boldmath $\ell^+\ell^-\gamma$ Final States}
\label{sss:llg}

\begin{figure}[b]
\centerline{\hbox{
\epsfig{file=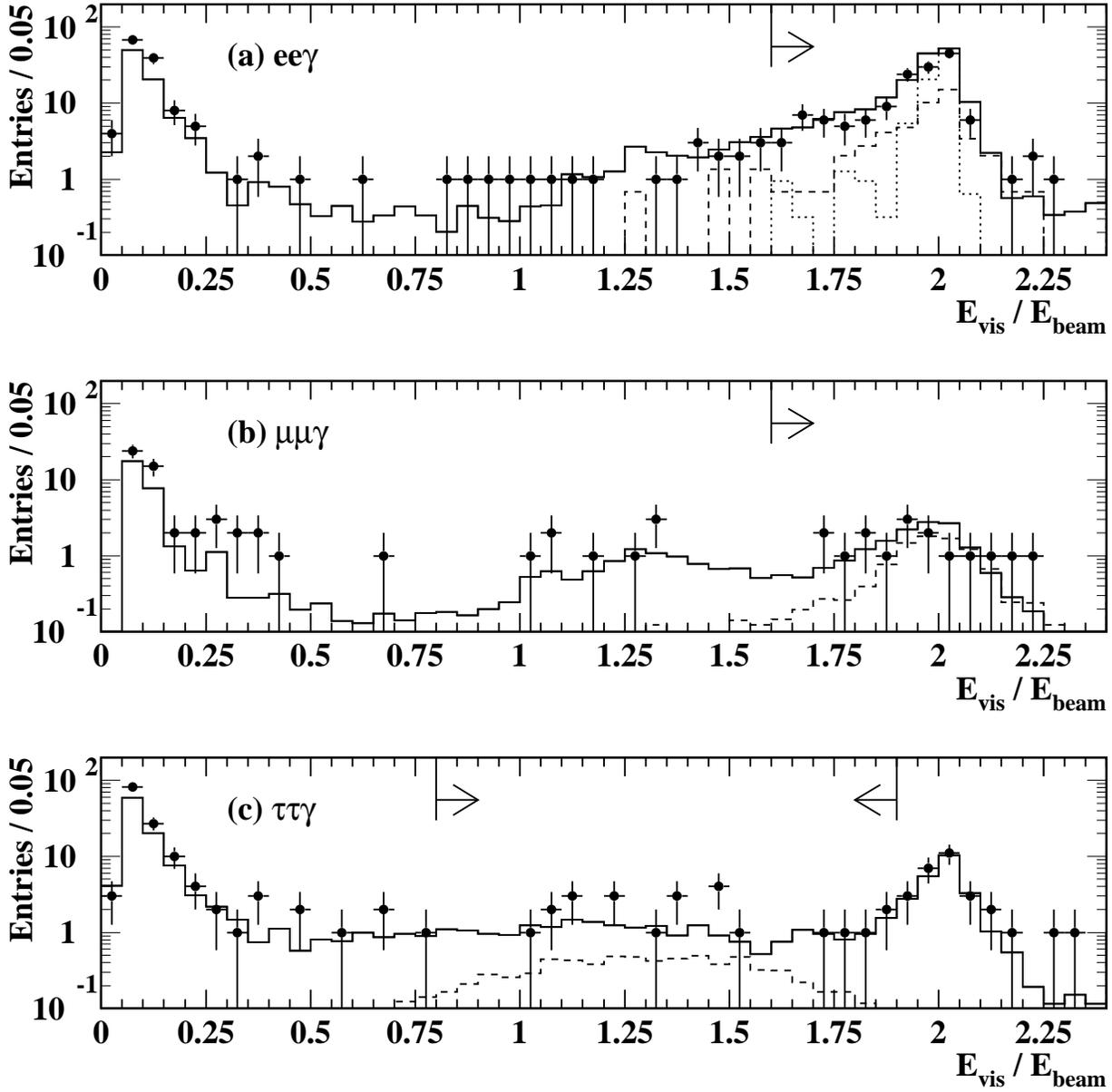,width=\textwidth,
  bbllx=25pt,bblly=150pt,bburx=540pt,bbury=680pt}
}}
\caption{
  The sums of the energies of the two most
  energetic leptons and most energetic photon,
  after requiring at least two leptons of the same
  flavour and one photon in the event.
  (a) is for ${\rm e^+e^-}\gamma$,
  (b) is for $\mu^+\mu^-\gamma$   and
  (c) is for $\tau^+\tau^-\gamma$.
  The dashed and dotted
  lines are signal Monte Carlo with
  $f/\Lambda=$(200~GeV)$^{-1}$
  and with an $\ell^{*\pm}$ mass of 140 GeV and 170~GeV, respectively,
  the solid lines are the sum of all of the Standard Model background
  Monte Carlo and the filled
  circles are the data.
  The arrows indicate the cut positions.
}
\label{f:llgen}
\end{figure}

Candidate $\ell^+\ell^-\gamma$ events are required to satisfy the
following criteria:

\begin{enumerate}
  \item There must be at least two identified leptons of the same
        flavour and at least one photon.
        The two most energetic leptons and
        the most energetic photon are used for further analysis.
        For the excited tau search, it is required that at most one
        of the two tau jets be an identified electron, and at most one
        of the two tau jets be an identified muon.
  \item In the ${\rm e^+e^-}\gamma$ and $\mu^+\mu^-\gamma$
        analyses, the sum of the energies of the two leptons and one photon,
        $E_{\rm vis}$, must satisfy
        $E_{\rm vis} > 1.6~E_{\rm beam}$.
        In the $\tau^+\tau^-\gamma$
        analysis, the sum of the energies of the two tau jets
        and one photon must satisfy
        $0.8~E_{\rm beam} < E_{\rm vis} < 1.9~E_{\rm beam}$.
        The sums of the energies of the two leptons
        and the photon before cut~2 are plotted in Figure~\ref{f:llgen}.
  \item A significant background after cuts~1 and 2 is from Bhabha scattering
        or lepton-pair production with final-state radiation.
        This is reduced by requiring the lepton-photon pair to
        be isolated.  The minimum opening angle between both
        lepton-photon combinations, $\theta_{\rm min}^{\ell\gamma}$
        is required to satisfy
        $|\cos\theta_{\rm min}^{\ell\gamma}|<$~0.90 for
        ${\rm e^+e^-}\gamma$ and $|\cos\theta_{\rm min}^{\ell\gamma}|<$~0.95 
        for $\mu^+\mu^-\gamma$ and $\tau^+\tau^-\gamma$.
        For the ${\rm e}^*$ search, radiative Bhabha scattering is
        further suppressed by also requiring that 
        the photon and at least one electron satisfy
        $|\cos{\theta}|<0.7$.
  \item The momenta of the two leptons and one photon are now
        calculated assuming a 3-body final state
        using the measured angles of the tracks
        and clusters.  The calculation uses the beam-energy
        constraint, conserving energy and momentum.
        One initial-state radiation photon along the beam
        axis is included in the calculation.
        It is required that all energies calculated are greater
        than zero.  The momenta of the leptons and photon calculated in
        this manner are used in cut~5, and also to construct the
        lepton-photon invariant mass for the excited lepton search.
  \item The final reducible component in the background
        is from the radiative-return process
        ${\rm e^+e^-}\rightarrow {\rm Z^0}\gamma
        \rightarrow \ell^+\ell^-\gamma$.
        Events with a di-lepton mass,
        $M_{\ell\ell}$, satisfying
        80~GeV~$<M_{\ell\ell}<$~100~GeV are vetoed.
\end{enumerate}

The complete analysis is summarized in Table~\ref{t:ssum},
including two example mass points for single-production of
charged excited leptons with photonic decays.
The numbers of events selected in each channel are
consistent with the expectations from Standard Model
sources.

\begin{figure}[b]
\centerline{\hbox{
\epsfig{file=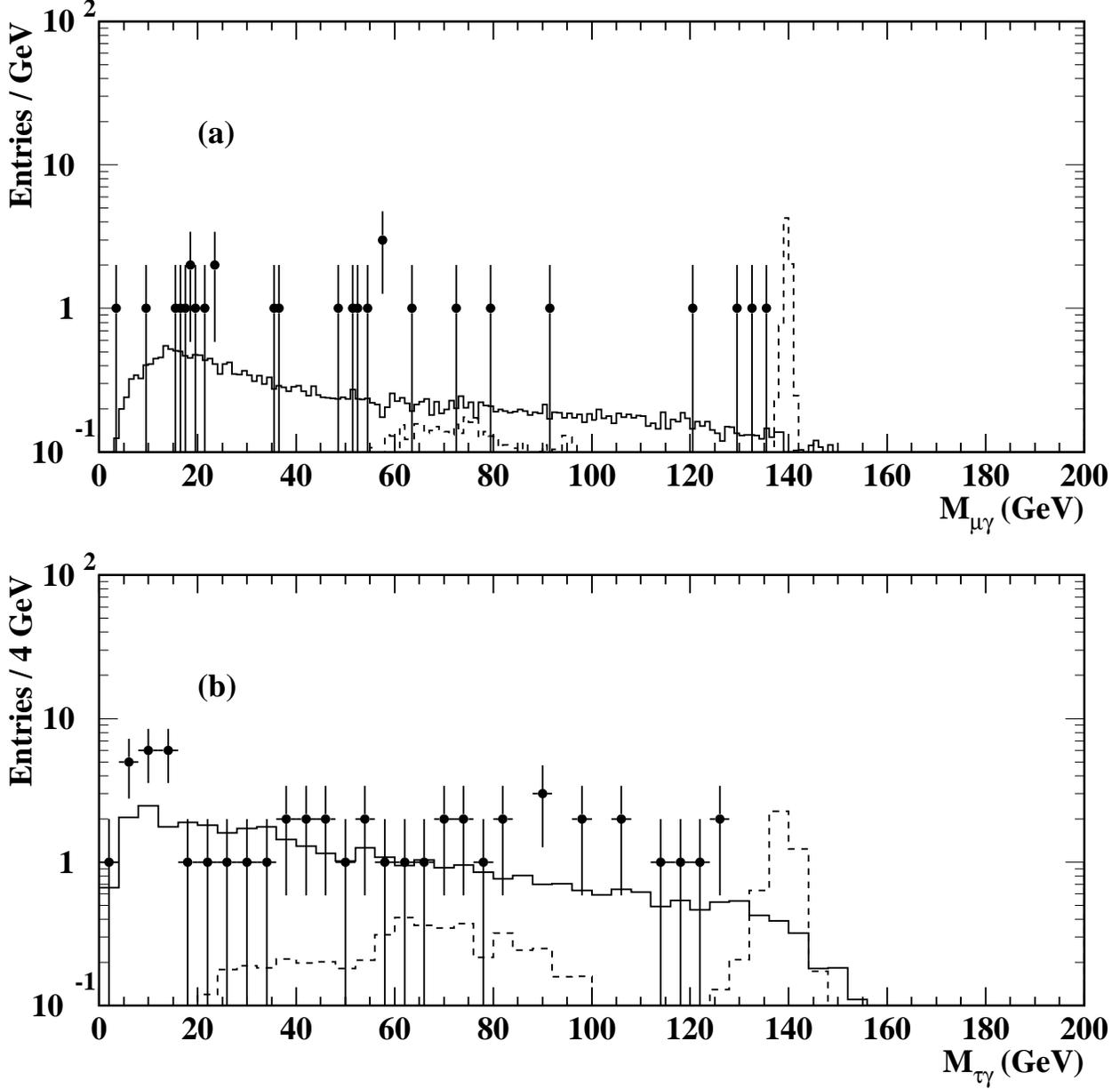,width=\textwidth,
  bbllx=25pt,bblly=150pt,bburx=540pt,bbury=680pt}
}}
\caption{
  Single production of charged excited leptons with photonic decays:
  $\ell^\pm\gamma$
  invariant mass distributions after all cuts for
  the full 161--172~GeV data set.
  (a) is $\mu^+\mu^-\gamma$ and (b) is
  $\tau^+\tau^-\gamma$.
  The dashed
  lines are signal Monte Carlo with $f/\Lambda=$(100~GeV)$^{-1}$
  and with an $\ell^{*\pm}$ mass of 140 GeV,
  the solid lines are the sum of all of the Standard Model background
  Monte Carlo and the filled
  circles are the data.  There are two entries per event.
}
\label{f:masslg}
\end{figure}

The lepton-photon mass resolutions are estimated with Monte Carlo
to be 0.4--0.5~GeV for excited electrons and muons with photonic decays,
and 2--3~GeV for excited taus with photonic decays.
The lepton-photon invariant masses, $M_{\ell\gamma}$, are plotted in
Figure~\ref{f:masslg} for the excited muon and tau searches, 
combining the full 161--172~GeV data.
No anomalous structure is apparent.
When computing limits for an excited lepton with mass $M_*$,
all events with at least one lepton-photon mass combination
satisfying
$|M_{\ell\gamma}-M_*|<\Delta$ are considered signal
candidates, where $\Delta=$~2 and 5~GeV for
the excited muon and tau search, respectively.
The efficiency for observing the single-production of excited
leptons with photonic decays is estimated with Monte Carlo
to be about 70\% for $\mu^*\mu$ and 40\% for $\tau^*\tau$,
including the mass window constraint,
with some dependence
on the mass of the excited lepton.
The excited electron results are summarized after
the analysis described in Section~\ref{sss:eg}.

\subsubsection{\boldmath ${\rm e}^*$ Single Production: ${\rm e}\gamma$ Topology}
\label{sss:eg}

A significant fraction of
singly-produced ${\rm e}^*$ events would have the recoil electron at a
small polar angle, outside the detector acceptance,
making the search for the ${\rm e}\gamma$ final state
also interesting.

After removing events which pass the ${\rm ee}\gamma$
selection described in Section~\ref{sss:llg},
candidate ${\rm e}\gamma$ events are required to satisfy the
following criteria:

\begin{enumerate}
  \item There must be at least one identified electron
        and one identified photon.  The most energetic
        electron and photon are used for further analysis.
  \item The energy sum of the electron and photon, $E_{\rm vis}$,
        must satisfy $E_{\rm vis} > 0.8~E_{\rm beam}$.
  \item A significant background after cuts~1 and 2 is from
        Bhabha scattering
        with final-state radiation.
        This is reduced by requiring the electron-photon pair to
        be isolated.  The opening angle between the
        electron and photon, $\theta_{\rm min}^{\ell\gamma}$,
        is required to satisfy
        $|\cos\theta_{\rm min}^{\ell\gamma}|<$~0.95.
  \item Bhabha scattering is further suppressed by requiring
        the photon to satisfy $|\cos{\theta_\gamma}|<0.7$,
        and that the event thrust axis formed with the
        observed lepton and photon satisfy
        $|\cos{\theta_{\rm thrust}}|<0.9$.
  \item The momenta of the electron and photon are now calculated
        assuming a 2-body final state plus one electron missing
        along the beam axis using only the measured polar angles
        of the observed electron and photon.
        The calculation uses the beam-energy constraint, conserving
        energy and momentum.  It is required that all energies
        calculated are greater than zero.  The momenta of the electron
        and photon calculated in this manner are used to construct
        the electron-photon invariant mass for the excited
        electron search.
\end{enumerate}

The analysis is summarized in Table~\ref{t:ssum},
including two example mass points for single-production of
charged excited electrons with photonic decays.
The number of events selected is
consistent with the expectation from Standard Model
sources.

\begin{figure}[b]
\centerline{\hbox{
\epsfig{file=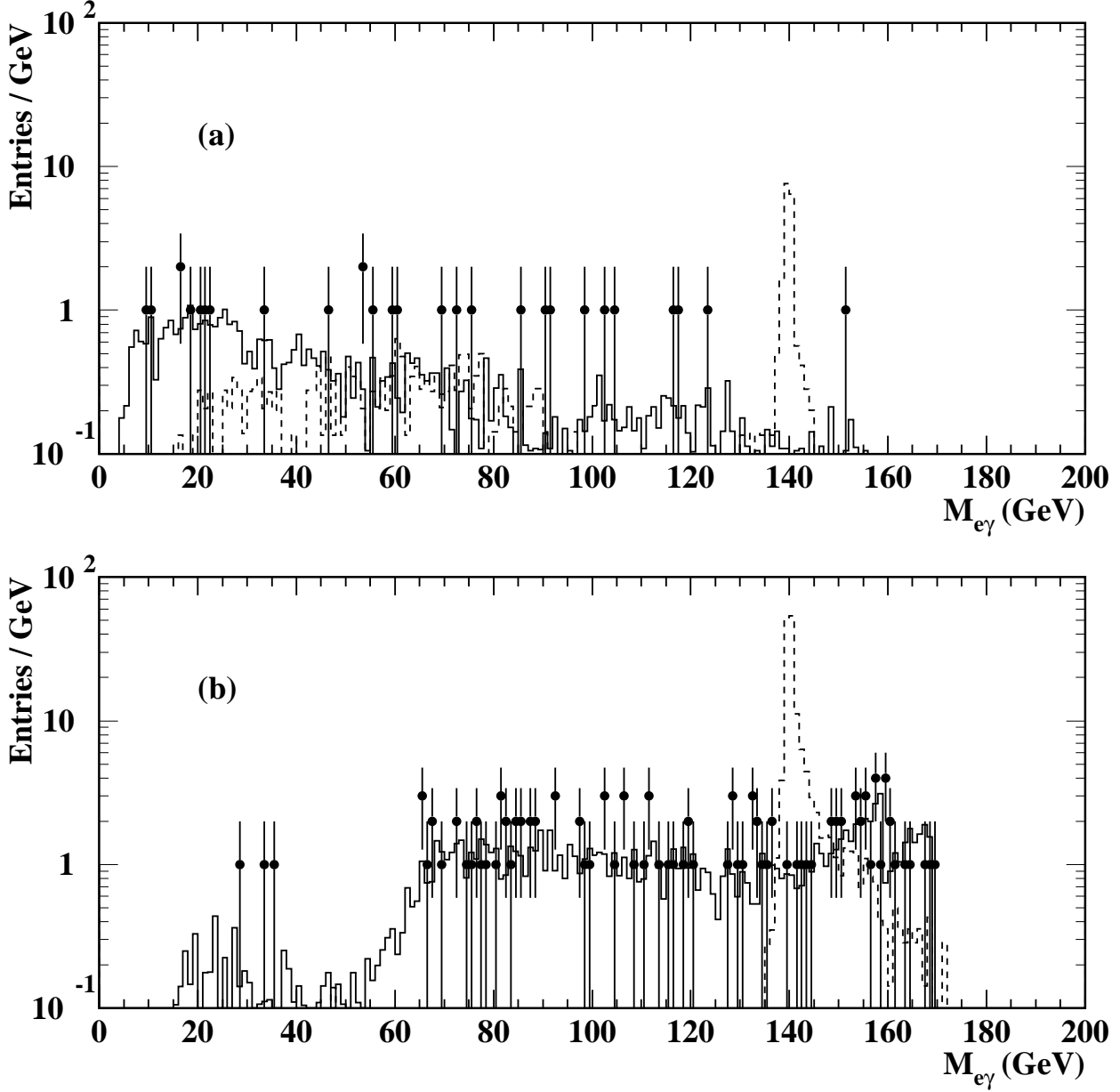,width=\textwidth,
  bbllx=25pt,bblly=150pt,bburx=540pt,bbury=680pt}
}}
\caption{
  Single production of charged excited electrons with photonic decays:
  ${\rm e}\gamma$
  invariant mass distributions after all cuts for
  the full 161--172~GeV data set.
  (a) The ${\rm e^+e^-}\gamma$ analysis and
  (b) the ${\rm e}\gamma$ (missing electron) topology.
  The dashed
  lines are signal Monte Carlo with $f/\Lambda=$(500~GeV)$^{-1}$
  and with an ${\rm e}^{*\pm}$ mass of 140 GeV,
  the solid lines are the sum of all of the Standard Model background
  Monte Carlo and the filled
  circles are the data.  (a) has two entries per event.
}
\label{f:masseg}
\end{figure}

Events which survive in either the ${\rm e^+e^-\gamma}$ or
${\rm e\gamma}$ analysis are considered as excited
electron candidates.
The electron-photon mass resolutions are estimated with Monte Carlo
to be 0.4--0.5~GeV in both channels.
The electron-photon invariant masses are plotted in
Figure~\ref{f:masseg},
combining the full 161--172~GeV data.
No anomalous structure is apparent.
When computing limits for an excited electron with mass $M_*$,
events satisfying
$|M_{\rm e \gamma}-M_*|<$~2~GeV are considered signal
candidates.
The efficiency for observing the single-production of excited
electrons with photonic decays is estimated with Monte Carlo
to be 35--70\%, including the mass window constraint,
depending on the mass of the excited electron.

\begin{table}
\begin{center}
\begin{tabular}{|c|c||c|c||c|c|c||c|c|} \hline
                      & After & Data & Total  &  ee$(\gamma)$ & $\mu\mu(\gamma)$ & $\tau\tau(\gamma)$ & $M_*=$ & $M_*=$ \\
                        & Cut &      &  Bkg   &        &          &            & 80 GeV & 140 GeV \\ \hline
                        &  1  &  320 &  310.9 &  219.4 &      0.0 &       4.0  &   7.5    &  11.0     \\
                        &  2  &  171 &  198.3 &  197.6 &      0.0 &       0.1  &   7.0    &  10.9     \\
 ${\rm ee}\gamma$       &  3  &   10 &   12.4 &   12.3 &      0.0 &       0.0  &   3.4    &   7.1     \\
                        &  4  &   10 &   12.4 &   12.3 &      0.0 &       0.0  &   3.4    &   7.0     \\
                        &  5  &    6 &   10.9 &   10.9 &      0.0 &       0.0  &   2.8    &   5.9     \\ \hline
                        &  1  & 3046 & 2384.5 & 1135.4 &      0.0 &      15.9  &  74.4    &  84.6     \\
                        &  2  &  669 &  722.1 &  714.3 &      0.0 &       4.2  &  73.3    &  84.5     \\
 ${\rm e}\gamma$        &  3  &  620 &  669.0 &  662.2 &      0.0 &       3.6  &  73.2    &  84.0     \\
                        &  4  &   61 &   81.1 &   78.7 &      0.0 &       1.8  &  42.2    &  62.9     \\
                        &  5  &   52 &   73.0 &   70.8 &      0.0 &       1.6  &  42.2    &  62.4     \\ \hline
                        &  1  &   76 &   55.9 &    0.0 &     23.3 &       2.5  &  85.0    &  84.6     \\
                        &  2  &   16 &   15.6 &    0.0 &     15.5 &       0.1  &  78.6    &  83.9     \\
 $\mu\mu\gamma$         &  3  &   11 &   12.5 &    0.0 &     12.3 &       0.0  &  77.7    &  81.5     \\
                        &  4  &   11 &   12.3 &    0.0 &     12.2 &       0.0  &  77.6    &  81.2     \\
                        &  5  &    9 &    8.1 &    0.0 &      8.1 &       0.0  &  70.5    &  67.7     \\ \hline
                        &  1  &  193 &  156.1 &   34.9 &      0.4 &      20.0  &  62.7    &  56.8     \\
                        &  2  &   23 &   23.4 &    8.1 &      0.3 &      13.2  &  57.1    &  54.1     \\
 $\tau\tau\gamma$       &  3  &   13 &   14.1 &    3.9 &      0.2 &       8.7  &  56.0    &  52.5     \\
                        &  4  &   13 &   12.6 &    3.4 &      0.2 &       7.9  &  54.9    &  51.4     \\
                        &  5  &    8 &    9.7 &    3.2 &      0.1 &       5.4  &  50.5    &  43.2     \\ \hline
\end{tabular}
\end{center}
\caption{
  The number of events surviving each cut for background,
  and the selection efficiencies in percent for
  excited lepton signal Monte Carlo, in the single-production search.
  The signal Monte Carlo corresponds to the coupling
  assignment $f=f^\prime$, but with 100\% photonic decays.
  The expected background levels are 
  normalized to the actual integrated luminosity.
  The efficiencies for the ${\rm e^+e^-\gamma}$ and 
  ${\rm e\gamma}$ excited electron single-production
  topologies are exclusive, and can be summed for the
  total selection efficiency.
  The statistical error on the background Monte Carlo is small
  compared to that for data.
  }
\label{t:ssum}
\end{table}


\section{Results}
\label{s:results}

The numbers of expected signal events are evaluated from the total
production cross-sections, the total integrated luminosity, and
the estimated detection efficiencies for the various analyses.
In the pair-production searches, the production cross-section
is relatively model-independent, and limits on the masses of
the heavy or excited lepton can be obtained directly.  In the
single-production searches, the production cross-section
depends on parameters within the model, so limits on those
parameters, as a function of the new particle masses, are inferred
instead.  The systematic errors on the total number of expected
signal events are estimated from:
     the statistical error on the
     Monte Carlo estimate of the detection efficiency, 1--2\%;
     the error due to the interpolation used to infer the
     efficiency at arbitrary masses from the limited number
     of Monte Carlo samples, 2--5\%;
     the error on the integrated luminosity, 0.6\%; the
     uncertainties in modelling the lepton identification
     cuts are estimated by comparing the efficiency for selecting
     leptons in events from Standard Model processes between
     data and Monte Carlo, and are found to be in the range 2--5\%;
     the uncertainty in the modelling of the photon conversion
     finder used in the $\ell^+\ell^-\gamma(\gamma)$ analyses
     is estimated by comparing the ratio of non-converting to
     converting photons in data and Standard Model process
     Monte Carlo, and is found to be
     1\% per photon in the event.
The errors are considered to be independent and are added in
quadrature to give the total systematic error, which is always
less than 10\%.  The systematic error is incorporated into the
limits inferred from the data  using the method described in
Reference~\cite{ref:cousins}.  The final limits are computed
considering the background levels expected in the analysis \cite{ref:pdgbg},
including the full 161--172~GeV dataset.

\begin{figure}
\centerline{\hbox{
\epsfig{file=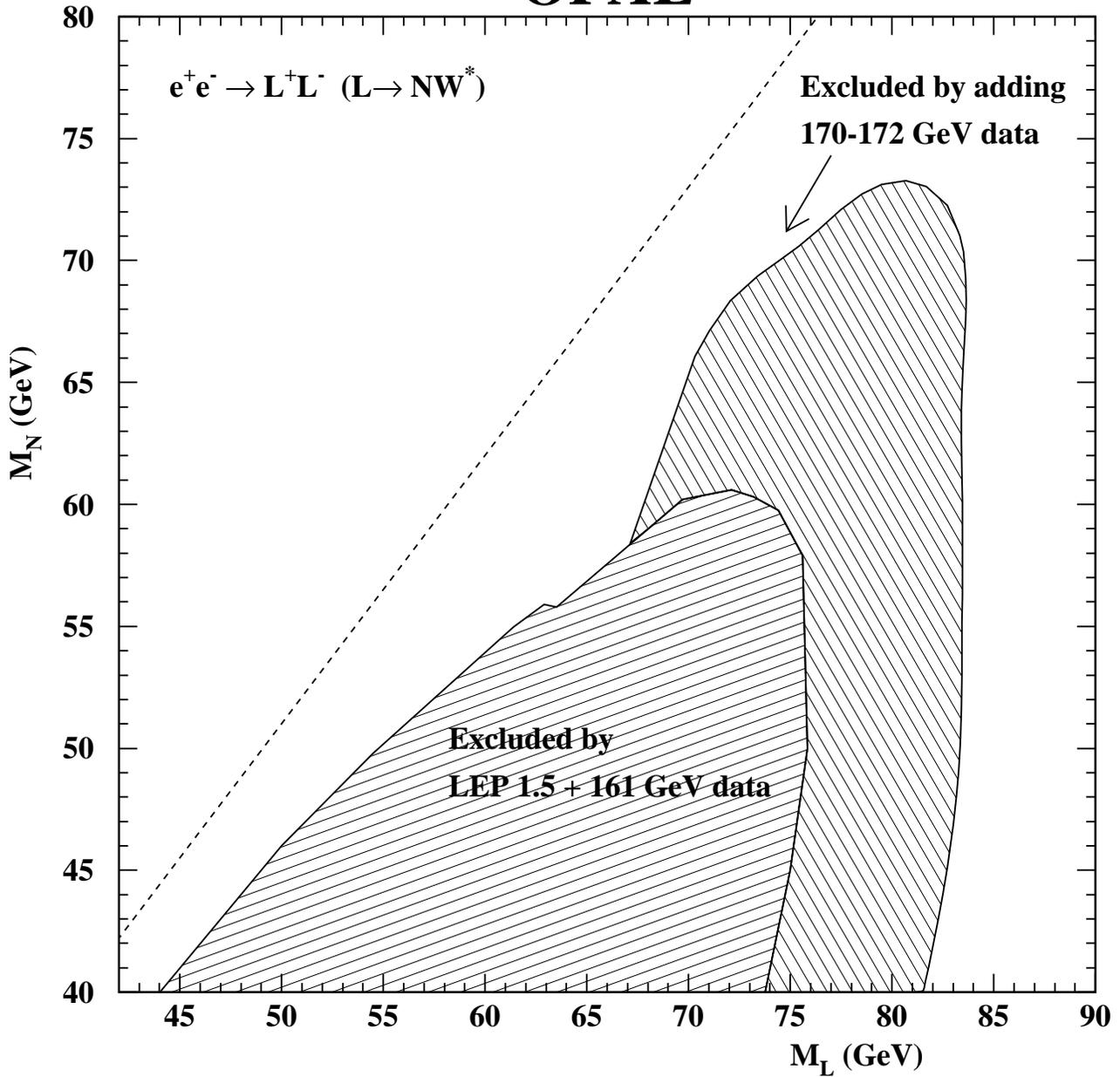,width=\textwidth,
  bbllx=0pt,bblly=0pt,bburx=520pt,bbury=550pt}
}}
\caption{Excluded region in the $(M_{\rm L}, M_{\rm N})$ plane for ${\rm L^+ L^-}$
    production in the case where ${\rm L^- \rightarrow N W^-}$.  The two 
    hatched regions indicated the previous 
    (LEP 1.5 at $\protect\sqrt{s}=$~130 and 136~GeV, and the 
    $\protect\sqrt{s}=$~161~GeV data)
    and new (170-172~GeV data) analyses.
  }
\label{f:hllim}
\end{figure}

For the flavour-mixing heavy lepton decays in which the
pair-produced heavy leptons undergo CC decays into
light leptons, 95\% confidence level lower limits on the mass
of the heavy lepton are shown in Table~\ref{t:hlmasslim}.
These results are valid for a mixing angle squared, $\zeta^2$,
greater than about 10$^{-12}$.
In the case that a heavy charged lepton, ${\rm L^\pm}$, decays into a
stable heavy neutral lepton, N, the exclusion region
depends on both $M_{\rm L}$ and $M_{\rm N}$.  The excluded
region is shown in Figure~\ref{f:hllim}.  From that figure,
if the charged heavy lepton is at least 8.4~GeV more massive
than the neutral heavy lepton, its mass is less than 81.5~GeV
at the 95\% confidence level.

\begin{table}
\begin{center}
\begin{tabular}{|c|c|c|}\hline
   \multicolumn{2}{|c|}{Mode}                             & Mass Limit                       \\
    \multicolumn{2}{|c|}{}                                &  (GeV)                           \\ \hline\hline
${\rm N \rightarrow e W}$           & Dirac               &  79.1                            \\
                                    & Majorana            &  69.8                            \\ \hline
${\rm N \rightarrow \mu W}$         & Dirac               &  78.5                            \\
                                    & Majorana            &  68.7                            \\ \hline
${\rm N \rightarrow \tau W}$        & Dirac               &  69.0                            \\
                                    & Majorana            &  54.4                            \\ \hline
\multicolumn{2}{|c|}{${\rm L^-\rightarrow \nu_\ell W^-}$} &  80.2                            \\ \hline
\multicolumn{2}{|c|}{${\rm L^-\rightarrow N W^-}$}        &  81.5 (with $\Delta M > 8.4$)  \\
\hline
\end{tabular}
\caption{95\% confidence level lower
  mass limits on unstable neutral heavy leptons
  obtained from the combined data
  collected at $\protect\sqrt{s}=$~161, 170 and 172~GeV.}
\label{t:hlmasslim}
\end{center}
\end{table}

\begin{figure}
\centerline{\hbox{
\epsfig{file=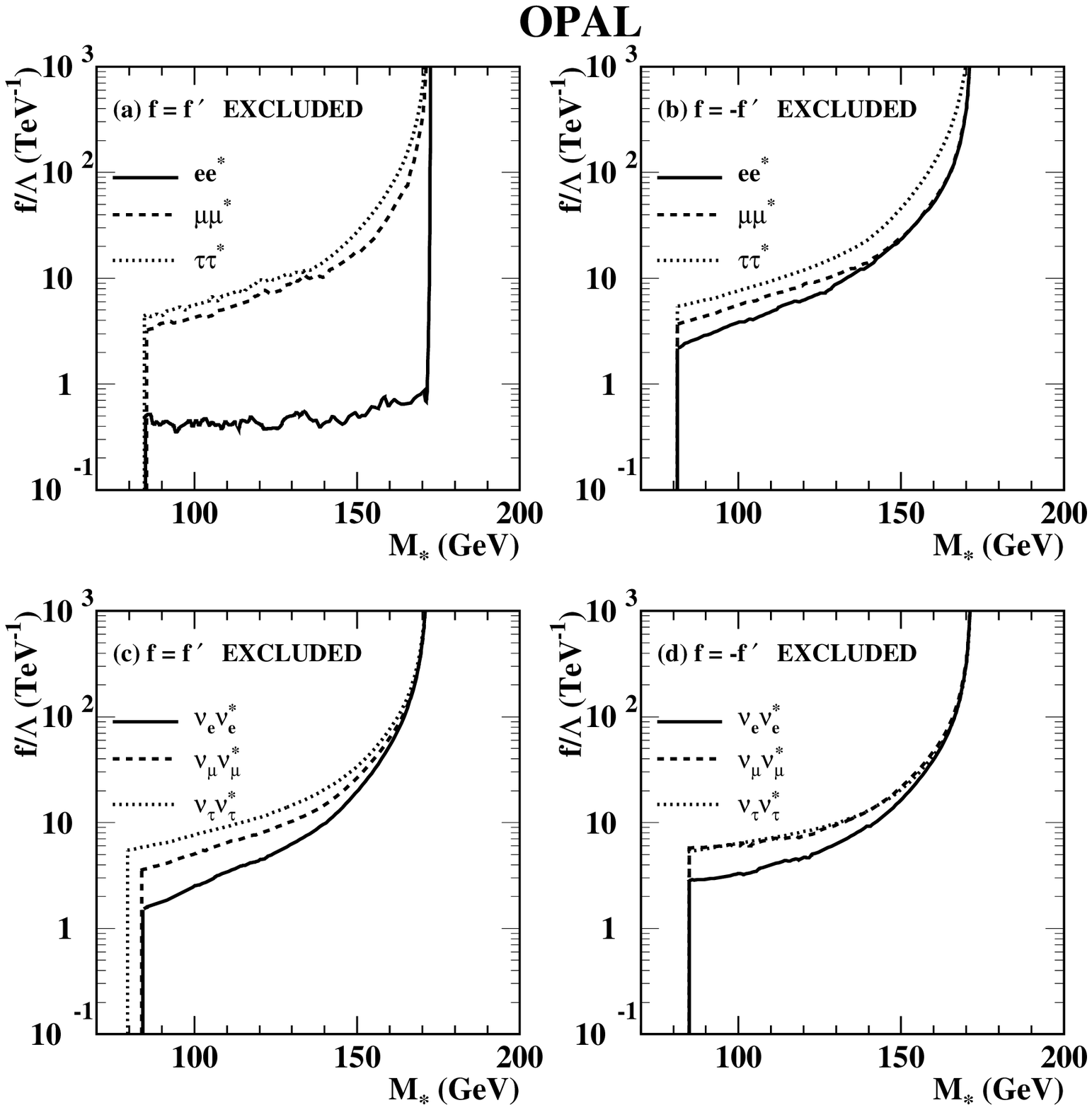,width=\textwidth,
  bbllx=25pt,bblly=160pt,bburx=535pt,bbury=680pt}
}}
\caption{95\% confidence level upper limits on the ratio of the coupling
  to the compositeness scale,
  $f/\Lambda$, as a function of the
  excited lepton mass.  (a) shows the limits on ${\rm e}^*$,
  $\mu^*$ and $\tau^*$ with $f=f^\prime$,
  (b) shows the limits on  ${\rm e}^*$, $\mu^*$ and $\tau^*$ with $f=-f^\prime$,
  (c) shows the limits on $\nu_{\rm e}^*$, $\nu_\mu^*$ and $\nu_\tau^*$ with
  $f=f^\prime$,
  and (d) shows the
  limits on $\nu_{\rm e}^*$, $\nu_\mu^*$ and $\nu_\tau^*$ with
  $f=-f^\prime$.
  The regions above and to the left of the
  curves are excluded by the single- and pair-production searches,
  respectively.
  }
\label{f:elcouplim}
\end{figure}

The mass limits on excited leptons are somewhat better than
for the heavy lepton case, primarily due to the vector couplings
leading to larger production cross-sections.  The mass
limits inferred from the pair-production searches
are shown in Table~\ref{t:elmasslim}.  Since the branching
ratio to either photons or W-bosons is essentially 100\%
in the mass range that
our pair-production searches are sensitive to, only the dominant decay
is used when computing the mass limits.
From the single-production
searches, limits on the ratio of the coupling to the compositeness
scale, $f/\Lambda$, are shown in Figure~\ref{f:elcouplim} for
two example coupling assumptions.  Since the branching ratio of
the excited lepton decays via the different vector bosons is 
arbitrary, example coupling assignments $f=\pm f^\prime$
are used to calculate these branching ratios and then the
photonic and CC decay
results are combined for the limits.

\begin{table}
\begin{center}
\begin{tabular}{|c|c|l|c|} \hline
   Flavour         &   Coupling    & Dominant      & Mass Limit \\
                   &               & Decay         & (GeV)      \\ \hline\hline
   ${\rm e}^*$     & $f=f^\prime$  & Photonic      & 85.0       \\
   $\mu^*$         & $f=f^\prime$  & Photonic      & 85.3       \\
   $\tau^*$        & $f=f^\prime$  & Photonic      & 84.6       \\ \hline
   ${\rm e}^*$     & $f=-f^\prime$ & Charged       & 81.3       \\
   $\mu^*$         & $f=-f^\prime$ & Charged       & 81.3       \\
   $\tau^*$        & $f=-f^\prime$ & Charged       & 81.3       \\ \hline   
   $\nu_{\rm e}^*$ & $f=f^\prime$  & Charged       & 84.1       \\
   $\nu_\mu^*$     & $f=f^\prime$  & Charged       & 83.9       \\
   $\nu_\tau^*$    & $f=f^\prime$  & Charged       & 79.4       \\ \hline
   $\nu_{\rm e}^*$ & $f=-f^\prime$ & Photonic      & 84.9       \\
   $\nu_\mu^*$     & $f=-f^\prime$ & Photonic      & 84.9       \\
   $\nu_\tau^*$    & $f=-f^\prime$ & Photonic      & 84.9       \\ \hline
\end{tabular}
\end{center}
\caption{ 95\% confidence level lower
  mass limits for the different excited leptons obtained from the
  pair-production searches.  The coupling assumption affects
  the branching ratio.
  }
\label{t:elmasslim}
\end{table}


\section{Conclusion}
\label{s:conclusion}

We have analysed a data sample corresponding to an integrated luminosity of
$10.3~{\rm pb}^{-1}$ at 170.3 and 172.3~GeV,
collected with the OPAL detector at LEP, to search for the
production of unstable heavy and excited leptons.
No evidence for their existence
was found, and limits on masses and couplings were established.
From the search for the pair production of heavy and excited leptons,
lower mass limits were established.
From the search for the single production of excited leptons,
upper limits on the ratio
of the coupling to the compositeness scale were derived.
These limits substantially improve those from previous
LEP searches.


\section{Acknowledgements}
\label{s:acknow}

We particularly wish to thank the SL Division for the efficient operation
of the LEP accelerator at all energies
and for
their continuing close cooperation with
our experimental group.  We thank our colleagues from CEA, DAPNIA/SPP,
CE-Saclay for their efforts over the years on the time-of-flight and trigger
systems which we continue to use.  In addition to the support staff at our own
institutions we are pleased to acknowledge the  \\
Department of Energy, USA, \\
National Science Foundation, USA, \\
Particle Physics and Astronomy Research Council, UK, \\
Natural Sciences and Engineering Research Council, Canada, \\
Israel Science Foundation, administered by the Israel
Academy of Science and Humanities, \\
Minerva Gesellschaft, \\
Benoziyo Center for High Energy Physics,\\
Japanese Ministry of Education, Science and Culture (the
Monbusho) and a grant under the Monbusho International
Science Research Program,\\
German Israeli Bi-national Science Foundation (GIF), \\
Bundesministerium f\"ur Bildung, Wissenschaft,
Forschung und Technologie, Germany, \\
National Research Council of Canada, \\
Hungarian Foundation for Scientific Research, OTKA T-016660, 
T023793 and OTKA F-023259.\\


\newpage

\newpage

\newpage


\newpage


\newpage


\newpage


\newpage


\newpage


\newpage


\newpage


\newpage



\begin{thebibliography}{99}

\bibitem{ref:pdg} ``Review of Particle Physics''
R.M. Barnett {\it et al.}, Phys. Rev. D54 (1996).
  
\bibitem{ref:revue}
A.~Djouadi, D.~Schaile, C.~Verzegnassi, {\it et al.}, Report of the Working
Group ``Extended Gauge Models'' in
Proceedings of the Workshop ``${\rm e^+e^-}$ Collisions at 500 GeV: The Physics 
Potential'', P.~Zerwas, (ed.) Report DESY 92-123A+B.

\bibitem{ref:hllep1}
ALEPH Collaboration, D.~Decamp   {\it et al.}, Phys. Lett. B236 (1990) 511; \\
OPAL Collaboration, M.Z.~Akrawy  {\it et al.}, Phys. Lett. B240 (1990) 250; \\
OPAL Collaboration, M.Z.~Akrawy  {\it et al.}, Phys. Lett. B247 (1990) 448; \\
L3 Collaboration, B.~Adeva       {\it et al.}, Phys. Lett. B251 (1990) 321; \\
DELPHI Collaboration, P.~Abreu   {\it et al.}, Phys. Lett. B274 (1992) 230.

\bibitem{ref:hlOPAL15}
OPAL Collaboration, G. Alexander {\it et al.}, Phys. Lett. B385 (1996) 433.

\bibitem{ref:hllep15}
L3 Collaboration,    M.~Acciarri {\it et al.}, Phys. Lett. B377 (1996) 304;  \\
ALEPH Collaboration, D.~Buskulic {\it et al.}, Phys. Lett. B384 (1996) 439.

\bibitem{ref:hlOPAL161}
  OPAL Collaboration, K.~Ackerstaff {\it et al.}, Phys. Lett. B393 (1997) 217.

\bibitem{ref:hlL3172}
  L3 Collaboration, M.~Acciarri {\it et al.}, 
  ``Search for Heavy Neutral and Charged Leptons in ${\rm e^+e^-}$ Annihilation
    at $\protect\sqrt{s}=$~161 and 172~GeV'', CERN-PPE/97-75, submitted to
    Phys. Lett. B.

\bibitem{ref:ellep1}
OPAL Collaboration, M.Z.~Akrawy {\it et al.}, Phys. Lett. B257 (1990) 531;\\
ALEPH Collaboration, D.~Decamp {\it et al.},   Phys. Lett. B250 (1990) 172;\\
DELPHI Collaboration, P.~Abreu {\it et al.},   Z.~Phys.    C53  (1992) 41; \\
L3 Collaboration, M.~Acciarri {\it et al.},    Phys. Lett. B353 (1995) 136.

\bibitem{ref:elopal15}
OPAL Collaboration, G.~Alexander {\it et al.}, Phys. Lett. B386 (1996) 463.

\bibitem{ref:ellep15}
L3 Collaboration, M.~Acciarri {\it et al.},   Phys. Lett. B370 (1996) 211;\\
DELPHI Collaboration, P.~Abreu {\it et al.},  Phys. Lett. B380 (1996) 480;\\
ALEPH Collaboration, D.~Buskulic {\it et al.},Phys. Lett. B385 (1996) 445.

\bibitem{ref:elopal161}
  OPAL Collaboration, K.~Ackerstaff {\it et al.}, Phys. Lett. B391 (1997) 197.
  
\bibitem{ref:ellep161}
  DELPHI Collaboration, P.~Abreu {\it et al.}, Phys. Lett. B393 (1997) 245;\\
  L3 Collaboration, M. Acciarri {\it et al.},  Phys. Lett. B401 (1997) 139.
  
\bibitem{ref:herasearches}
H1 Collaboration, I.~Abt {\it et al.}, Nucl. Phys. B396 (1993) 3;\\
ZEUS Collaboration, M.~Derrick {\it et al.}, Z. Phys. C65 (1994) 627.

\bibitem{ref:opalgg}
OPAL Collaboration, G.~Alexander {\it et al.}, Phys. Lett. B377 (1996) 222.

\bibitem{ref:2f} OPAL Collaboration, K.~Ackerstaff {\it et al.}, Phys. Lett. B391 (1997) 221;\\
  OPAL Collaboration, K.~Ackerstaff {\it et al.}, ``Tests of the Standard Model and Constraints 
  on New Physics from Measurements of Fermion-pair Production at 130-172 GeV at LEP'',
  CERN-PPE/97-101, submitted to Z. Phys C.

\bibitem{ref:abdel} A. Djouadi, Z. Phys. C63 (1994) 317, and references therein.

\bibitem{ref:nardi} E. Nardi, E. Roulet and D. Tommasini, Phys. Lett. B344 (1995) 225.

\bibitem{ref:bdk} F.~Boudjema, A.~Djouadi and J.L.~Kneur, Z.~Phys. C57 (1993) 425.

\bibitem{ref:OPAL-detector}
OPAL Collaboration, K.~Ahmet {\it et al.}, Nucl.~Instr.~Meth. A305 (1991) 275;\\
P.P.~Allport {\it et al.}, Nucl.~Instr.~Meth. A324 (1993) 34;\\
P.P.~Allport {\it et al.}, Nucl.~Instr.~Meth. A346 (1994) 476;\\
B.E.~Anderson {\it et al.}, IEEE Trans.~Nucl.~Sci. 41 (1994) 845.

\bibitem{ref:zerwas} J.H.~K\"{u}hn, A.~Reiter and P.M.~Zerwas, Nucl. Phys. B272 (1986) 560.

\bibitem{ref:jetset} T. Sj\"{o}strand, Comp.~Phys.~Comm. 82 (1994) 74.

\bibitem{ref:bhwide}
S.~Jadach, W.~Placzek and B.F.L.~Ward, Phys. Lett. B390 (1997) 298.

\bibitem{ref:teegg} D.~Karlen, Nucl. Phys. B289 (1987) 23.

\bibitem{ref:koralz}
S.~Jadach, B.F.L.~Ward and Z.~W\c{a}s, Comp.~Phys.~Comm. 79 (1994) 503.

\bibitem{ref:herwig} G.~Marchesini {\it et al.}, Comp.~Phys.~Comm. 67 (1992) 465.

\bibitem{ref:phojet} A.~Buijs {\it el al.}, Phys. Rev. D54 (1996) 4244.

\bibitem{ref:vermaseren}
R.~Bhattacharya, J.~Smith and G.~Grammer, Phys. Rev. D15 (1977) 3267;\\
J.~Smith, J.A.M.~Vermaseren and G.~Grammer, Phys. Rev. D15 (1977) 3280.

\bibitem{ref:fermisv}
  J.~Hilgart, R.Kleiss and F.~Le Diberder, Comp.~Phys.~Comm. 75 (1993) 191.

\bibitem{ref:excalibur}
 F.A.~Berends, R.~Pittau and R.~Kleiss, Comp. Phys. Comm. 85 (1995) 437.

\bibitem{ref:grc4f} 
  J.~Fujimoto {\it et al.}, Comp. Phys. Comm. 100 (1996) 128.

\bibitem{ref:radcor}
F.A.~Berends and R.~Kleiss, Nucl. Phys. B186 (1981) 22.

\bibitem{ref:nunugpv}
  G. Montagna {\it et al.}, Nucl. Phys. B452 (1995) 161.

\bibitem{ref:gopal}
J.~Allison {\it et al.}, Nucl.~Instr.~Meth. A317 (1992) 47.

\bibitem{ref:gce}
  OPAL Collaboration, M.Z.~Akrawy {\it el al.}, Phys. Lett. B253 (1990) 511.

\bibitem{ref:durham} 
  N.~Brown and W.J.~Stirling, Phys. Lett. B252 (1990) 657; \\
  S.~Bethke, Z.~Kunszt, D.~Soper and W.J.~Stirling, Nucl. Phys. B370 (1992) 310; \\
  S.~Catani {\it et al.}, Nucl. Phys. B269 (1991) 432; \\
  N.~Brown and W.~J.~Stirling, Z. Phys. C53 (1992) 629.

\bibitem{ref:NN}
  OPAL Collaboration, R.~Akers {\it et al.}, Phys. Lett. B327 (1994) 411.
  
\bibitem{ref:ggOPAL172}
  OPAL Collaboration, K.~Ackerstaff {\it et al.},
  ``Search for Anomalous Production of Photonic Events with Missing Energy
    in ${\rm e^+e^-}$ Collisions at $\sqrt{s}=$~130-172~GeV'',
    to be submitted to Z. Phys. C.
  
\bibitem{ref:LL} OPAL Collaboration, G.~Alexander {\it et al.}, Z.~Phys. C52 (1991) 175.

\bibitem{ref:dedx} M. Hauschild {\it et al.}, Nucl. Instr. and Meth. A314 (1992) 74.

\bibitem{ref:MU}
  OPAL Collaboration, P. Acton {\it et al.},
  Z.~Phys. C58 (1993) 523.

\bibitem{ref:cousins} R.D.~Cousins and V.L.~Highland, 
  Nucl.~Instr.~Meth. A 320 (1992) 331.

\bibitem{ref:pdgbg} Formulae 28.40 in Reference~\cite{ref:pdg}.
  
\end{thebibliography}
\end{document}